# Alternative Derivation of Electromagnetic Cloaks and Concentrators

**A. D. Yaghjian**[1] **and S. Maci**[2]

[1] AFRL/RYHA, Hanscom AFB, MA 01731, USA
[2] University of Siena, Siena 53100, Italy
E-mail: a.yaghjian@verizon.net and macis@dii.unisi.it

**Abstract:** Beginning with a straightforward formulation of electromagnetic 'cloaking' that reduces to a boundary value problem involving a single Maxwell first-order differential equation, explicit formulae for the relative permittivity–permeability dyadic and fields of spherical and circular cylindrical annular cloaks are derived in terms of general compressed radial coordinate functions. The general formulation is based on the requirements that the cloaking occurs for all possible incident fields and that the cloaks with frequency $\omega > 0$ have continuous tangential components of **E** and **H** fields across their outer surfaces, and zero normal components of **D** and **B** fields at their inner material surfaces. The tangential-field boundary conditions at the outer surface of the cloak ensure zero scattered fields, and the normal-field boundary conditions at the inner surface of the cloak are compatible with zero total fields inside the interior cavity of the cloak. For spherical cloaks, unlike cylindrical cloaks, these boundary conditions lead to all the tangential components of the **E** and **H** fields being continuously zero across their inner surfaces—cylindrical cloaks having delta functions in polarization densities at their inner surfaces. For *H*-wave incident fields, a nonmagnetic circular cylindrical annulus is found that has nonzero scattered fields but zero total fields within its interior cavity. For bodies with no interior free-space cavities, the formulation is used to derive nonscattering spherical and cylindrical 'concentrators' that magnify the incident fields near their centers. For static fields ($\omega = 0$), the boundary value formulation is appropriately modified to obtain a relative permeability dyadic that will cloak magnetostatic fields. Causality-energy conditions imply that, unlike magnetostatic cloaking, electrostatic cloaking as well as low-frequency cloaking for $\omega > 0$ is not realizable. It is also confirmed that perfect cloaking over any nonzero bandwidth violates causality-energy conditions and thus the cloaking of realistic time-dependent fields must be approximate.

## 1. Introduction

This work began after reading the paper by Pendry, Schurig, and Smith [1] on electromagnetic cloaking and realizing that the material parameters and fields of specific cloaks could be derived as a boundary value problem with a single first-order Maxwell differential equation for linear anisotropic media without relying heavily on coordinate transformations [2]–[5]. This alternative formulation of cloaking reveals the boundary values of the fields at the inner and outer surfaces of a cloak that yield zero scattered fields outside the cloak and zero total fields inside the free-space cavity of the cloak. The coordinate transformation functions and constitutive parameters (permittivity and permeability) are exhibited in a form that is frequency independent for frequency $\omega > 0$ and thus cloaking is immediately extendable (ignoring causality-energy considerations; see section 6) to time-domain fields through Fourier transformation of the frequency-domain fields. The boundary value formulation confirms the important discovery of Greenleaf *et al* [6] that there are delta functions in the tangential polarization densities at the inner surfaces of cylindrical cloaks. The boundary value formulation is also applied (in section 5) to obtain the coordinate transformation functions and constitutive parameters that produce cloaking of static fields ($\omega = 0$). The boundary value formulation presented here is intended to complement the 'transformation optics' of [1] and the 'conformal mapping' of [7], each of which provide a conceptually appealing approach to the design of complex electromagnetic structures. The formulation given in [1] is based on spatial coordinate transformations and corresponding transformations of Maxwell's equations that provide expressions for the required inhomogeneity and anisotropy of the permittivity and permeability of the cloaking material. An approximate cloaking method was presented by Leonhardt [7] where the Helmholtz equation is transformed by conformal mapping to produce cloaking effects in the geometric optics limit. Subsequently, Leonhardt and Philbin discussed transformation media in the context of the transformation rules of tensors in general relativity [4]. Conformal mapping is also applied in elastodynamics [5].

Numerical studies [8]–[13] have confirmed the possibility of electromagnetic cloaking either by using commercial finite-element equation solvers [12, 13] or by employing the decomposition of the electromagnetic fields into a set of orthogonal eigenmodes [8]–[10]. Simplified cloaks are proposed in [11, 13] that use approximations to the anisotropic tensors of an ideal cloak.

Several different kinds of 'cloaking' can be found in the published literature in addition of the cloaking introduced by Pendry *et al* [1]. Kerker *et al* [14, 15] found that coated spheres and ellipsoids could be made electromagnetically invisible to plane waves in the long wavelength regime. Alu and Engheta [16]–[19] showed that electrically small metallic spheres, like electrically small antennas [20, 21] can be made nearly transparent to the impinging radiation.

Coating to minimize the forward scattering from a cylinder was introduced in [22] and [23]. Kildal *et al* showed in [24] how cylinders such as struts and masts can be constructed with reduced blockage widths in order to decrease the sidelobes and losses caused by the blockage of reflector antenna fields. Skokic *et al* [25] showed that a metallic parabolic reflector is invisible to a monostatic radar under certain resonant conditions.

A cloaking technique via a change of variables for electric impedance tomography was introduced by Greenleaf *et al* [26] and developed further in a recent paper by Kohn *et al* [27].

Milton *et al* [28] show that the scattering vanishes from a fixed distribution of a finite number of polarizable dipoles within a near-field region of a cylindrical 'superlens'.

Miller [29] proposed the use of sensors and active sources near a closed surface to cloak the region inside the surface from illumination by an arbitrary time-dependent incident electromagnetic wave.

## 2. Formulation

We begin the boundary value formulation with the following form of Maxwell's equations for $\exp(-i\omega t)$, $\omega > 0$, time dependence

$$\nabla \times \mathbf{E}(\mathbf{r}) - i\omega \bar{\boldsymbol{\mu}}(\mathbf{r}) \cdot \mathbf{H}(\mathbf{r}) = -\mathbf{J}_{\mathrm{m}}^{\mathrm{inc}}(\mathbf{r}), \tag{1a}$$

$$\nabla \times \mathbf{H}(\mathbf{r}) + i\omega \bar{\boldsymbol{\epsilon}}(\mathbf{r}) \cdot \mathbf{E}(\mathbf{r}) = \mathbf{J}_{\mathrm{e}}^{\mathrm{inc}}(\mathbf{r}) \tag{1b}$$

with

$$\mathbf{D}(\mathbf{r}) = \bar{\boldsymbol{\epsilon}}(\mathbf{r}) \cdot \mathbf{E}(\mathbf{r}), \tag{2a}$$

$$\mathbf{B}(\mathbf{r}) = \bar{\boldsymbol{\mu}}(\mathbf{r}) \cdot \mathbf{H}(\mathbf{r}) \tag{2b}$$

where **E** and **H** are the electric and magnetic fields (**D** is the electric displacement and **B** is the magnetic induction), $\bar{\boldsymbol{\epsilon}}$ and $\bar{\boldsymbol{\mu}}$ are the permittivity and permeability dyadics, and $(\mathbf{J}_{\mathrm{e}}^{\mathrm{inc}}, \mathbf{J}_{\mathrm{m}}^{\mathrm{inc}})$ are the electric and magnetic source current densities of the incident fields $[\mathbf{E}^{\mathrm{inc}}(\mathbf{r}), \mathbf{H}^{\mathrm{inc}}(\mathbf{r})]$ illuminating the electromagnetic cloak from outside the cloak. The incident fields satisfy the Maxwell equations

$$\nabla \times \mathbf{E}^{\text{inc}}(\mathbf{r}) - i\omega\mu_0 \mathbf{H}^{\text{inc}}(\mathbf{r}) = -\mathbf{J}_m^{\text{inc}}(\mathbf{r}), \tag{3a}$$

$$\nabla \times \mathbf{H}^{\text{inc}}(\mathbf{r}) + i\omega\epsilon_0 \mathbf{E}^{\text{inc}}(\mathbf{r}) = \mathbf{J}_e^{\text{inc}}(\mathbf{r}). \tag{3b}$$

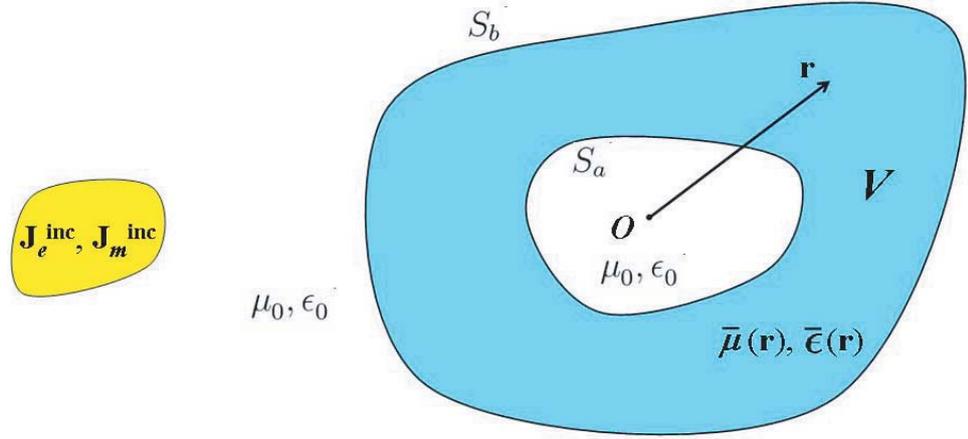

**Figure 1.** Cross-sectional geometry of cloak with sources of incident field.

The electromagnetic cloak is a general annular volume of material $V$ bounded by the interior surface $S_a$ and the exterior surface $S_b$ as shown in figure 1. The volume inside $S_a$ and outside $S_b$ is assumed to be free space with permittivity and permeability denoted by $\epsilon_0$ and $\mu_0$, that is, $\bar{\epsilon} = \epsilon_0 \bar{\mathbf{I}}$ and $\bar{\mu} = \mu_0 \bar{\mathbf{I}}$, where $\bar{\mathbf{I}}$ is the unit dyadic. The vector $\mathbf{r}$ is the position vector measured from a chosen origin $O$.

An effective electromagnetic cloak at the angular frequency $\omega$ produces zero scattered fields outside its exterior surface $S_b$ and zero total fields inside its interior surface $S_a$ for all incident fields $[\mathbf{E}^{\text{inc}}(\mathbf{r}), \mathbf{H}^{\text{inc}}(\mathbf{r})]$. Therefore, within the volume $V$ of the cloaking material, assume the electric and magnetic fields take the form

$$\mathbf{E}(\mathbf{r}) = \bar{\mathbf{A}}_e(\mathbf{r}) \cdot \mathbf{E}^{\text{inc}}[\mathbf{f}(\mathbf{r})], \tag{4a}$$

$$\mathbf{H}(\mathbf{r}) = \bar{\mathbf{A}}_h(\mathbf{r}) \cdot \mathbf{H}^{\text{inc}}[\mathbf{f}(\mathbf{r})], \tag{4b}$$

where $\bar{\mathbf{A}}_e(\mathbf{r})$ and $\bar{\mathbf{A}}_h(\mathbf{r})$ are dyadic functions of position $\mathbf{r}$ and $\mathbf{f}(\mathbf{r})$ is a real-valued vector coordinate transformation function of position ($\mathbf{r}$), which is shorthand notation for $(u, v, w)$, where $u, v$ and $w$ are the given coordinates of a particular three-dimensional coordinate system that conveniently represents the geometry of the cloak (such as $(u, v, w) = (r, \theta, \phi)$ for a spherical cloak). (The functions $\bar{\mathbf{A}}_e(\mathbf{r})$, $\bar{\mathbf{A}}_h(\mathbf{r})$, and $\mathbf{f}(\mathbf{r})$ are assumed independent of the values of the incident fields.) The vector function $[\mathbf{f}(\mathbf{r})]$ is shorthand notation for three given real-valued scalar functions $[f(u, v, w), g(u, v, w), h(u, v, w)]$. Thus, for example, $\mathbf{E}^{\text{inc}}(\mathbf{r})$ is shorthand notation for $\mathbf{E}^{\text{inc}}(u, v, w)$. The expression $\mathbf{E}^{\text{inc}}[\mathbf{f}(\mathbf{r})]$ is shorthand notation for $\mathbf{E}^{\text{inc}}[u \Rightarrow f(u, v, w), v \Rightarrow g(u, v, w), w \Rightarrow h(u, v, w)]$, where the symbol '$\Rightarrow$' means 'replaced by.' If $\mathbf{E}^{\text{inc}}(\mathbf{r})$ is written in terms of its components along the unit vectors $(\hat{\mathbf{u}}, \hat{\mathbf{v}}, \hat{\mathbf{w}})$, that is,

$$\mathbf{E}^{\text{inc}}(\mathbf{r}) = E_1^{\text{inc}}(u, v, w)\hat{\mathbf{u}} + E_2^{\text{inc}}(u, v, w)\hat{\mathbf{v}} + E_3^{\text{inc}}(u, v, w)\hat{\mathbf{w}}, \tag{5a}$$

then

$$\begin{aligned}\mathbf{E}^{\text{inc}}[\mathbf{f}(\mathbf{r})] = &\, E_1^{\text{inc}}[u \Rightarrow f(u,v,w), v \Rightarrow g(u,v,w), w \Rightarrow h(u,v,w)]\hat{\mathbf{u}} \\ &+ E_2^{\text{inc}}[u \Rightarrow f(u,v,w), v \Rightarrow g(u,v,w), w \Rightarrow h(u,v,w)]\hat{\mathbf{v}} \\ &+ E_3^{\text{inc}}[u \Rightarrow f(u,v,w), v \Rightarrow g(u,v,w), w \Rightarrow h(u,v,w)]\hat{\mathbf{w}}. \end{aligned} \quad (5b)$$

In the free-space region outside $S_b$, we want the total fields to be just the incident fields, and in the free-space region inside the surface $S_a$, we want the total fields to be zero. This can be expressed mathematically as

$$\bar{\mathbf{A}}_e(\mathbf{r}) = \bar{\mathbf{A}}_h(\mathbf{r}) = \begin{cases} \bar{\mathbf{I}}, & \mathbf{r} \text{ outside } S_b, \\ 0, & \mathbf{r} \text{ inside } S_a, \end{cases} \quad (6a)$$

$$\mathbf{f}(\mathbf{r}) = \mathbf{r}, \quad \mathbf{r} \text{ outside } S_b \text{ and inside } S_a \quad (6b)$$

with the function $\mathbf{f}(\mathbf{r})$ continuous across $S_b$

$$\mathbf{f}(\mathbf{r} \to S_b^-) = \mathbf{r}, \quad (6c)$$

that is

$$\lim_{\mathbf{r} \to S_b^-} f(u,v,w) = u, \quad \lim_{\mathbf{r} \to S_b^-} g(u,v,w) = v, \quad \lim_{\mathbf{r} \to S_b^-} h(u,v,w) = w, \quad (6d)$$

where $\mathbf{r} \to S_b^-$ means $\mathbf{r}$ approaching $S_b$ from inside $S_b$. The total electric and magnetic fields everywhere can therefore be written as

$$\mathbf{E}(\mathbf{r}) = \begin{cases} \mathbf{E}^{\text{inc}}(\mathbf{r}), & \mathbf{r} \text{ outside } S_b, \\ \bar{\mathbf{A}}_e(\mathbf{r}) \cdot \mathbf{E}^{\text{inc}}[\mathbf{f}(\mathbf{r})], & \mathbf{r} \in V, \\ 0, & \mathbf{r} \text{ inside } S_a, \end{cases} \quad (7a)$$

$$\mathbf{H}(\mathbf{r}) = \begin{cases} \mathbf{H}^{\text{inc}}(\mathbf{r}), & \mathbf{r} \text{ outside } S_b, \\ \bar{\mathbf{A}}_h(\mathbf{r}) \cdot \mathbf{H}^{\text{inc}}[\mathbf{f}(\mathbf{r})], & \mathbf{r} \in V, \\ 0, & \mathbf{r} \text{ inside } S_a. \end{cases} \quad (7b)$$

We can now determine the boundary conditions that should be applied at the outer and inner surfaces of the cloak. To this end, first note that a perfectly nonscattering cloak cannot absorb power and thus the material of the cloak must be lossless. Because electric and magnetic volume and surface charges and currents are zero in lossless material, Maxwell's equations (1) imply that the tangential components of $\mathbf{E}(\mathbf{r})$ and $\mathbf{H}(\mathbf{r})$ (and the normal components of $\mathbf{B}(\mathbf{r})$ and $\mathbf{D}(\mathbf{r})$) are continuous across the outer surface $S_b$ of the cloak unless there are delta functions in the fields at $S_b$. Since (6b) and (6c) express that the compression of the incident fields in (7) vanishes as the outer surface $S_b$ is approached from inside $S_b$, we can assume that there will be no delta functions in the fields at $S_b$ and thus the tangential components of $\mathbf{E}(\mathbf{r})$ and $\mathbf{H}(\mathbf{r})$ (and the normal components of $\mathbf{B}(\mathbf{r})$ and $\mathbf{D}(\mathbf{r})$) across $S_b$ will be continuous under the condition that $\mathbf{f}(\mathbf{r})$ satisfies (6b) and (6c). This continuity combines with (7) to yield

$$\hat{\mathbf{n}} \times \mathbf{E}^{\text{inc}}(\mathbf{r} \to S_b^+) = \hat{\mathbf{n}} \times \mathbf{E}^{\text{inc}}(\mathbf{r} \to S_b^-) = \hat{\mathbf{n}} \times \bar{\mathbf{A}}_e(\mathbf{r} \to S_b^-) \cdot \mathbf{E}^{\text{inc}}(\mathbf{r} \to S_b^-), \quad (8a)$$

$$\hat{\mathbf{n}} \times \mathbf{H}^{\text{inc}}(\mathbf{r} \to S_b^+) = \hat{\mathbf{n}} \times \mathbf{H}^{\text{inc}}(\mathbf{r} \to S_b^-) = \hat{\mathbf{n}} \times \bar{\mathbf{A}}_h(\mathbf{r} \to S_b^-) \cdot \mathbf{H}^{\text{inc}}(\mathbf{r} \to S_b^-), \quad (8b)$$

where $\hat{\mathbf{n}}$ is the unit normal to $S_b$ and $\mathbf{r} \to S_b^{-(+)}$ means $\mathbf{r}$ approaching $S_b$ from inside (outside) $S_b$. The second equations in (8a) and (8b) are equivalent to merely

$$\hat{\mathbf{n}} \times \bar{\mathbf{A}}_e(\mathbf{r}) \stackrel{\mathbf{r} \to S_b^-}{=} \hat{\mathbf{n}} \times \bar{\mathbf{A}}_h(\mathbf{r}) \stackrel{\mathbf{r} \to S_b^-}{=} \hat{\mathbf{n}} \times \bar{\mathbf{I}}. \tag{9}$$

The boundary conditions on the fields as $\mathbf{r} \to S_a^+$ ($\mathbf{r}$ approaches $S_a$ from outside $S_a$) must be compatible with the fields inside $S_a$ being zero. Continuous tangential components of $\mathbf{E}(\mathbf{r})$ and $\mathbf{H}(\mathbf{r})$ fields across $S_a$ that are both zero as $\mathbf{r} \to S_a^+$ would imply from the Maxwell free-space equations inside $S_a$ that the fields will be zero inside $S_a$. However, demanding that all the tangential components of $\mathbf{E}(\mathbf{r})$ and $\mathbf{H}(\mathbf{r})$ be zero as $\mathbf{r} \to S_a^+$ may yield an overdetermined boundary value problem. Thus, we shall not demand that the tangential components of the $\mathbf{E}(\mathbf{r})$ and $\mathbf{H}(\mathbf{r})$ fields be zero as $\mathbf{r} \to S_a^+$. Instead, we shall impose the less restrictive boundary conditions that the normal components of the $\mathbf{B}(\mathbf{r})$ and $\mathbf{D}(\mathbf{r})$ fields both be zero as $\mathbf{r} \to S_a^+$; that is

$$\hat{\mathbf{n}} \cdot \mathbf{D}(\mathbf{r}) \stackrel{\mathbf{r} \to S_a^+}{=} \hat{\mathbf{n}} \cdot \left\{ \bar{\boldsymbol{\epsilon}}(\mathbf{r}) \cdot \bar{\mathbf{A}}_e(\mathbf{r}) \cdot \mathbf{E}^{\text{inc}}[\mathbf{f}(\mathbf{r})] \right\} \stackrel{\mathbf{r} \to S_a^+}{=} 0, \tag{10a}$$

$$\hat{\mathbf{n}} \cdot \mathbf{B}(\mathbf{r}) \stackrel{\mathbf{r} \to S_a^+}{=} \hat{\mathbf{n}} \cdot \left\{ \bar{\boldsymbol{\mu}}(\mathbf{r}) \cdot \bar{\mathbf{A}}_h(\mathbf{r}) \cdot \mathbf{H}^{\text{inc}}[\mathbf{f}(\mathbf{r})] \right\} \stackrel{\mathbf{r} \to S_a^+}{=} 0 \tag{10b}$$

or, because $\mathbf{E}^{\text{inc}}$ and $\mathbf{H}^{\text{inc}}$ are arbitrary, simply

$$\hat{\mathbf{n}} \cdot \left\{ \bar{\boldsymbol{\epsilon}}(\mathbf{r}) \cdot \bar{\mathbf{A}}_e(\mathbf{r}) \right\} \stackrel{\mathbf{r} \to S_a^+}{=} 0, \tag{10c}$$

$$\hat{\mathbf{n}} \cdot \left\{ \bar{\boldsymbol{\mu}}(\mathbf{r}) \cdot \bar{\mathbf{A}}_h(\mathbf{r}) \right\} \stackrel{\mathbf{r} \to S_a^+}{=} 0. \tag{10d}$$

If the fields are zero inside $S_a$, these boundary conditions in (10) imply that the normal components of $\mathbf{B}(\mathbf{r})$ and $\mathbf{D}(\mathbf{r})$ are continuously zero across $S_a$.

Substitution of $\mathbf{E}$ and $\mathbf{H}$ from (4) into (1) yields

$$\nabla \times \left[ \bar{\mathbf{A}}_e(\mathbf{r}) \cdot \mathbf{E}^{\text{inc}}[\mathbf{f}(\mathbf{r})] \right] - i\omega \bar{\boldsymbol{\mu}}(\mathbf{r}) \cdot \bar{\mathbf{A}}_h(\mathbf{r}) \cdot \mathbf{H}^{\text{inc}}[\mathbf{f}(\mathbf{r})] = 0, \quad \mathbf{r} \in V, \tag{11a}$$

$$\nabla \times \left[ \bar{\mathbf{A}}_h(\mathbf{r}) \cdot \mathbf{H}^{\text{inc}}[\mathbf{f}(\mathbf{r})] \right] + i\omega \bar{\boldsymbol{\epsilon}}(\mathbf{r}) \cdot \bar{\mathbf{A}}_e(\mathbf{r}) \cdot \mathbf{E}^{\text{inc}}[\mathbf{f}(\mathbf{r})] = 0, \quad \mathbf{r} \in V. \tag{11b}$$

With $\mathbf{H}^{\text{inc}}$ and $\mathbf{E}^{\text{inc}}$ inserted from (3a) and (3b) into (11a) and (11b), respectively, these equations become

$$\nabla \times \left[ \bar{\mathbf{A}}_e(\mathbf{r}) \cdot \mathbf{E}^{\text{inc}}[\mathbf{f}(\mathbf{r})] \right] - \mu_0^{-1} \bar{\boldsymbol{\mu}}(\mathbf{r}) \cdot \bar{\mathbf{A}}_h(\mathbf{r}) \cdot \left[ \nabla \times \mathbf{E}^{\text{inc}}(\mathbf{r}) \right]_{\mathbf{r} \Rightarrow \mathbf{f}(\mathbf{r})} = 0, \quad \mathbf{r} \in V, \tag{12a}$$

$$\nabla \times \left[ \bar{\mathbf{A}}_h(\mathbf{r}) \cdot \mathbf{H}^{\text{inc}}[\mathbf{f}(\mathbf{r})] \right] - \epsilon_0^{-1} \bar{\boldsymbol{\epsilon}}(\mathbf{r}) \cdot \bar{\mathbf{A}}_e(\mathbf{r}) \cdot \left[ \nabla \times \mathbf{H}^{\text{inc}}(\mathbf{r}) \right]_{\mathbf{r} \Rightarrow \mathbf{f}(\mathbf{r})} = 0, \quad \mathbf{r} \in V, \tag{12b}$$

where, for example, as (5b) would indicate

$$\begin{aligned}
\left[ \nabla \times \mathbf{E}^{\text{inc}}(\mathbf{r}) \right]_{\mathbf{r} \Rightarrow \mathbf{f}(\mathbf{r})} &= \left[ \nabla \times \mathbf{E}^{\text{inc}}(u, v, w) \right]_{u \Rightarrow f(u,v,w), v \Rightarrow g(u,v,w), w \Rightarrow h(u,v,w)} \\
&= \hat{\mathbf{u}} \left[ \left( \nabla \times \mathbf{E}^{\text{inc}}(u, v, w) \right)_1 \right]_{u \Rightarrow f(u,v,w), v \Rightarrow g(u,v,w), w \Rightarrow h(u,v,w)} \\
&\quad + \hat{\mathbf{v}} \left[ \left( \nabla \times \mathbf{E}^{\text{inc}}(u, v, w) \right)_2 \right]_{u \Rightarrow f(u,v,w), v \Rightarrow g(u,v,w), w \Rightarrow h(u,v,w)} \\
&\quad + \hat{\mathbf{w}} \left[ \left( \nabla \times \mathbf{E}^{\text{inc}}(u, v, w) \right)_3 \right]_{u \Rightarrow f(u,v,w), v \Rightarrow g(u,v,w), w \Rightarrow h(u,v,w)}.
\end{aligned} \tag{13}$$

Equations (12*a*) and (12*b*) have to hold for all possible incident electric and magnetic fields, respectively, including the case where electric and magnetic fields interchange their roles, provided they are Maxwellian. Thus, assuming a unique solution for $\bar{\mathbf{A}}_e$, $\bar{\mathbf{A}}_h$ and $\mu_0^{-1}\bar{\boldsymbol{\mu}}$ exists to (12*a*) for a given cloak geometry and transformation function $\mathbf{f}(\mathbf{r})$, then a unique solution for $\bar{\mathbf{A}}_h$, $\bar{\mathbf{A}}_e$ and $\epsilon_0^{-1}\bar{\boldsymbol{\epsilon}}$ exists to (12*b*) for a given cloak geometry and transformation function $\mathbf{f}(\mathbf{r})$. Alternatively, we can assume a unique solution to each of the second-order differential equations obtainable from these first-order differential equations for a given cloak geometry and transformation function $\mathbf{f}(\mathbf{r})$. In either case, it follows that

$$\bar{\mathbf{A}}_e(\mathbf{r}) = \bar{\mathbf{A}}_h(\mathbf{r}) = \bar{\mathbf{A}}(\mathbf{r}) \tag{14a}$$

and

$$\frac{\bar{\boldsymbol{\mu}}(\mathbf{r})}{\mu_0} = \frac{\bar{\boldsymbol{\epsilon}}(\mathbf{r})}{\epsilon_0} = \bar{\boldsymbol{\alpha}}(\mathbf{r}). \tag{14b}$$

Consequently, the two equations in (12) are equivalent to each other and to find $\bar{\mathbf{A}}(\mathbf{r})$ and the relative permittivity–permeability dyadic $\bar{\boldsymbol{\alpha}}(\mathbf{r})$ for a given cloak geometry and coordinate function $\mathbf{f}(\mathbf{r})$, we need only solve one first-order differential equation, say

$$\nabla \times \left[\bar{\mathbf{A}}(\mathbf{r}) \cdot \mathbf{E}^{\text{inc}}[\mathbf{f}(\mathbf{r})]\right] - \bar{\boldsymbol{\alpha}}(\mathbf{r}) \cdot \bar{\mathbf{A}}(\mathbf{r}) \cdot \left[\nabla \times \mathbf{E}^{\text{inc}}(\mathbf{r})\right]_{\mathbf{r}\Rightarrow\mathbf{f}(\mathbf{r})} = 0, \quad \mathbf{r} \in V, \tag{15}$$

which must hold for all possible $\mathbf{E}^{\text{inc}}(\mathbf{r})$ with $\mathbf{f}(\mathbf{r})$ and $\bar{\mathbf{A}}(\mathbf{r})$ satisfying the boundary conditions in (6*c*), (9) and (10), namely

$$\mathbf{f}(\mathbf{r}) \stackrel{\mathbf{r}\to S_b^-}{=} \mathbf{r}, \tag{16a}$$

$$\hat{\mathbf{n}} \times \bar{\mathbf{A}}(\mathbf{r}) \stackrel{\mathbf{r}\to S_b^-}{=} \hat{\mathbf{n}} \times \bar{\mathbf{I}}, \tag{16b}$$

$$\hat{\mathbf{n}} \cdot \{\bar{\boldsymbol{\alpha}}(\mathbf{r}) \cdot \bar{\mathbf{A}}(\mathbf{r})\} \stackrel{\mathbf{r}}{\to} S_a^+ = 0. \tag{16c}$$

There appears to be no reason why, in principle, the range of the coordinate function $\mathbf{f}(\mathbf{r})$ could not extend outside $S_b$, the outer surface of the cloak, provided the sources of the incident fields lie outside this range. That is, equations (15) and (16) allow $\mathbf{f}(\mathbf{r}) \in V_0$, where the surface $S_0$ of $V_0$ encloses the outer surface $S_b$ of the cloak, provided the sources of the incident fields lie outside of $V_0$. (If the sources of the incident fields were located between $S_0$ and $S_b$, the equation (15) would not hold because terms involving the incident sources would be required on the right-hand sides of (12*a*) and (12*b*), the equations from which (15) is obtained.) In other words, incident fields in a region of free space outside the cloak can be reproduced as $\bar{\mathbf{A}}(\mathbf{r}) \cdot \mathbf{E}^{\text{inc}}[\mathbf{f}(\mathbf{r})]$ in a region inside the cloaking material.

The equations (15), (16*b*) and (16*c*) are independent of frequency for $\omega > 0$. Therefore, if $\mathbf{f}(\mathbf{r})$ is chosen independent of frequency, the boundary value problem given in (15) and (16) produces values of $\bar{\boldsymbol{\alpha}}$ and $\bar{\mathbf{A}}$ that are independent of frequency. It should be noted, however, that for electrostatic fields ($\omega = 0$), the equation (15) degenerates to $\nabla \times [\bar{\mathbf{A}}(\mathbf{r}) \cdot \mathbf{E}^{\text{inc}}[\mathbf{f}(\mathbf{r})]] = 0$ ($\nabla \times [\bar{\mathbf{A}}(\mathbf{r}) \cdot \mathbf{H}^{\text{inc}}[\mathbf{f}(\mathbf{r})]] = 0$ for magnetostatic fields), which cannot be satisfied for all incident static fields. The formulation valid for the cloaking of static fields is given in section 5.

In sections 3 and 4, we shall solve (15) under the boundary conditions in (16) to determine the relative permittivity–permeability dyadic $[\bar{\boldsymbol{\alpha}}(\mathbf{r})]$ and the field dyadic $[\bar{\mathbf{A}}(\mathbf{r})]$ in terms of the transformation function [**f**(**r**)] for spherical and circular cylindrical cloaks. In addition, we determine these quantities for spherical and circular cylindrical 'concentrators' that magnify the incident fields near their centers. However, first we show that the solution to the boundary value problem in (15) and (16) leads to zero scattered fields outside the cloak and zero total fields in the free-space cavity of the cloak.

*2.1. Fields inside and outside the material annulus of the cloak*

The boundary condition (16*b*) requiring that the tangential components of **E** and **H** are continuous across the outer surface $S_b$ of the cloak and equal to the tangential components of $\mathbf{E}^{\text{inc}}$ and $\mathbf{H}^{\text{inc}}$ imply that the components of the scattered electric and magnetic fields, $\mathbf{E}^{\text{sc}}$ and $\mathbf{H}^{\text{sc}}$, tangential to $S_b$ are zero just outside $S_b$. Thus, Poynting's vector of the scattered fields normal to the surface $S_b$ is zero, implying from Poynting's theorem that the power radiated by the scattered fields and thus the far scattered fields are zero. A spherical wave expansion then shows that the scattered fields outside a sphere that circumscribes the cloak is zero. Using analyticity of the scattered fields in the free space outside the cloak proves that the scattered fields everywhere outside the surface $S_b$ of the cloak are zero.

The boundary conditions in (16*c*), requiring that the normal components of the **D** and **B** fields be zero as **r** approaches the inner surface $S_a$ from within the material of the cloak, do not imply by themselves that the total fields inside $S_a$ are zero, because perfectly electrically conducting (PEC) and perfectly magnetically conducting (PMC) cavity modes at their resonant frequencies can exist inside $S_a$ with delta functions in tangential polarization at $S_a^+$ [30].

Nonetheless, after we solve the system of equations in (15) and (16) say for the spherical annulus, we find a solution exists for $\bar{\boldsymbol{\alpha}} = \bar{\boldsymbol{\epsilon}}/\epsilon_0 = \bar{\boldsymbol{\mu}}/\mu_0$ and the fields in $r > a$ at every frequency for a given **f**(**r**) and furthermore that the solution is compatible with (though does not necessarily require) zero total fields inside $S_a$. Moreover, inserting a lossy material inside the cavity of the cloak eliminates all fields within the cavity because the boundary value problem in (15) and (16) outside the cavity is not changed and power dissipation inside the cavity is not compatible with zero scattered fields. Thus, the homogeneous solutions inside the cavity of the cloak at the cavity resonant frequencies are uncoupled from the solutions exterior to the cavity that are produced by sources outside the cloak. This implies that the fields within the cavity of the cloak illuminated by incident fields that are initially zero will not produce fields inside the cavity of the cloak. By reciprocity, sources inside the cavity will produce zero fields outside the cloak and also zero fields in the material of the cloak (assuming any fields that emanate into the material of the cloak would radiate outside the cloak). Thus, any dicontinuous tangential E and H fields across $S_a$ require from Maxwell's equations that delta functions in tangential polarization densities are produced at $S_a^+$ by the sources inside the cavity [30].

Similarly, assuming a solution exists for $\bar{\boldsymbol{\alpha}} = \bar{\boldsymbol{\epsilon}}/\epsilon_0 = \bar{\boldsymbol{\mu}}/\mu_0$ and the fields outside $S_a$ for a given **f**(**r**) in any shaped cloak, the fields inside $S_a$ for any shaped cloak remain zero if they are initially zero. (Although we have not carried out the details of the proof, existence of solution can presumably be proven by solving the equations (15) and (16) in curvilinear coordinates to obtain the general form of the solution found in [1, 4] using transformational methods.)

## 3. Spherical cloaks

A cloak consisting of a spherical annulus of anisotropic material with inner radius $a$ and outer radius $b$ is conveniently described by the spherical coordinates $(u, v, w) = (r, \theta, \phi)$. In order for the spherical annulus to behave as a perfectly nonscattering cloak for all possible directions and values of the incident fields, symmetry demands that

$$[\mathbf{f(r)}] = [f(r), g = \theta, h = \phi] \tag{17a}$$

with boundary condition (according to (16a))

$$f(b) = b \tag{17b}$$

and

$$\bar{\mathbf{A}}(\mathbf{r}) = A_r(r)\hat{\mathbf{r}}\hat{\mathbf{r}} + A_s(r)\left(\hat{\boldsymbol{\theta}}\hat{\boldsymbol{\theta}} + \hat{\boldsymbol{\phi}}\hat{\boldsymbol{\phi}}\right) \tag{18a}$$

with boundary condition (according to (16b))

$$A_s(b) = 1. \tag{18b}$$

Furthermore

$$\bar{\boldsymbol{\alpha}}(\mathbf{r}) = \alpha_r(r)\hat{\mathbf{r}}\hat{\mathbf{r}} + \alpha_s(r)\left(\hat{\boldsymbol{\theta}}\hat{\boldsymbol{\theta}} + \hat{\boldsymbol{\phi}}\hat{\boldsymbol{\phi}}\right), \tag{19}$$

so that

$$\bar{\boldsymbol{\alpha}}(\mathbf{r}) \cdot \bar{\mathbf{A}}(\mathbf{r}) = \alpha_r(r)A_r(r)\hat{\mathbf{r}}\hat{\mathbf{r}} + \alpha_s(r)A_s(r)\left(\hat{\boldsymbol{\theta}}\hat{\boldsymbol{\theta}} + \hat{\boldsymbol{\phi}}\hat{\boldsymbol{\phi}}\right) \tag{20a}$$

with boundary condition (according to (16c))

$$\alpha_r(a)A_r(a) = 0 \tag{20b}$$

and

$$\bar{\mathbf{A}}(\mathbf{r}) \cdot \mathbf{E}^{\text{inc}}[\mathbf{f(r)}] = A_r(r)E_r^{\text{inc}}[f(r), \theta, \phi]\hat{\mathbf{r}} + A_s(r)\left(E_\theta^{\text{inc}}[f(r), \theta, \phi]\hat{\boldsymbol{\theta}} + E_\phi^{\text{inc}}[f(r), \theta, \phi]\hat{\boldsymbol{\phi}}\right)$$
$$= [A_r(r) - A_s(r)]E_r^{\text{inc}}[f(r), \theta, \phi]\hat{\mathbf{r}} + A_s(r)\mathbf{E}^{\text{inc}}[f(r), \theta, \phi], \tag{21}$$

The curl of the incident field in spherical coordinates is given by

$$\nabla \times \mathbf{E}^{\text{inc}}(\mathbf{r}) = \frac{1}{r}\left(\frac{\partial E_\phi^{\text{inc}}}{\partial \theta} + \frac{E_\phi^{\text{inc}}}{\tan\theta} - \frac{1}{\sin\theta}\frac{\partial E_\theta^{\text{inc}}}{\partial \phi}\right)\hat{\mathbf{r}}$$
$$+ \left(\frac{1}{r\sin\theta}\frac{\partial E_r^{\text{inc}}}{\partial \phi} - \frac{\partial E_\phi^{\text{inc}}}{\partial r} - \frac{E_\phi^{\text{inc}}}{r}\right)\hat{\boldsymbol{\theta}}$$
$$+ \left(\frac{\partial E_\theta^{\text{inc}}}{\partial r} + \frac{E_\theta^{\text{inc}}}{r} - \frac{1}{r}\frac{\partial E_r^{\text{inc}}}{\partial \theta}\right)\hat{\boldsymbol{\phi}} \tag{22a}$$

and thus

$$\left[\nabla \times \mathbf{E}^{\text{inc}}(\mathbf{r})\right]_{\mathbf{r}\Rightarrow\mathbf{f(r)}} = \frac{1}{f(r)}\left(\frac{\partial E_\phi^{\text{inc}}}{\partial \theta} + \frac{E_\phi^{\text{inc}}}{\tan\theta} - \frac{1}{\sin\theta}\frac{\partial E_\theta^{\text{inc}}}{\partial \phi}\right)\hat{\mathbf{r}}$$
$$+ \left(\frac{1}{f(r)\sin\theta}\frac{\partial E_r^{\text{inc}}}{\partial \phi} - \frac{1}{f'(r)}\frac{\partial E_\phi^{\text{inc}}}{\partial r} - \frac{E_\phi^{\text{inc}}}{f(r)}\right)\hat{\boldsymbol{\theta}}$$
$$+ \left(\frac{1}{f'(r)}\frac{\partial E_\theta^{\text{inc}}}{\partial r} + \frac{E_\theta^{\text{inc}}}{f(r)} - \frac{1}{f(r)}\frac{\partial E_r^{\text{inc}}}{\partial \theta}\right)\hat{\boldsymbol{\phi}}, \tag{22b}$$

which can be rewritten as

$$\left[\nabla \times \mathbf{E}^{\text{inc}}(\mathbf{r})\right]_{\mathbf{r}\Rightarrow\mathbf{f}(\mathbf{r})} = \frac{r}{f(r)}\nabla \times \mathbf{E}^{\text{inc}}[f(r),\theta,\phi] + \left[\frac{1}{f'(r)} - \frac{r}{f(r)}\right]\hat{\mathbf{r}}$$
$$\times \frac{\partial \mathbf{E}^{\text{inc}}[f(r),\theta,\phi]}{\partial r}, \qquad (22c)$$

where $f'(r) = \mathrm{d}f(r)/\mathrm{d}r$. From (21) and (22a) we can write

$$\nabla \times \left[\bar{\mathbf{A}}(\mathbf{r}) \cdot \mathbf{E}^{\text{inc}}[\mathbf{f}(\mathbf{r})]\right] = A_s(r)\nabla \times \mathbf{E}^{\text{inc}}[f(r),\theta,\phi] + \frac{\mathrm{d}A_s(r)}{\mathrm{d}r}\hat{\mathbf{r}} \times \mathbf{E}^{\text{inc}}[f(r),\theta,\phi]$$
$$+ [A_s(r) - A_r(r)]\hat{\mathbf{r}} \times \nabla E_r^{\text{inc}}[f(r),\theta,\phi]. \qquad (23)$$

Inserting (20a), (22c) and (23) into (15), taking the radial and tangential components, and noting that

$$\left[\nabla \times \mathbf{E}^{\text{inc}}[f(r),\theta,\phi]\right]_{\text{tan}} = -\hat{\mathbf{r}} \times \nabla E_r^{\text{inc}}[f(r),\theta,\phi] + \frac{1}{r}\hat{\mathbf{r}} \times \mathbf{E}^{\text{inc}}[f(r),\theta,\phi]$$
$$+ \hat{\mathbf{r}} \times \frac{\partial \mathbf{E}^{\text{inc}}[f(r),\theta,\phi]}{\partial r} \qquad (24)$$

produces

$$\left[\frac{A_r(r)}{A_s(r)} - \frac{f(r)}{r\alpha_r(r)}\right]\left[\nabla \times \mathbf{E}^{\text{inc}}[f(r),\theta,\phi]\right]_r = 0, \qquad (25a)$$

$$-\left[\frac{A_r(r)}{A_s(r)} - \frac{r\alpha_s(r)}{f(r)}\right]\hat{\mathbf{r}} \times \nabla E_r^{\text{inc}}[f(r),\theta,\phi]$$
$$+ \left[1 - \frac{\alpha_s(r)}{f'(r)}\right]\hat{\mathbf{r}} \times \frac{\partial \mathbf{E}^{\text{inc}}[f(r),\theta,\phi]}{\partial r}$$
$$+ \left[\frac{1}{A_s(r)}\frac{\mathrm{d}A_s(r)}{\mathrm{d}r} + \frac{1}{r} - \frac{\alpha_s(r)}{f(r)}\right]\hat{\mathbf{r}} \times \mathbf{E}^{\text{inc}}[f(r),\theta,\phi] = 0. \qquad (25b)$$

Since the incident fields $\hat{\mathbf{r}} \times \nabla E_r^{\text{inc}}$, $\hat{\mathbf{r}} \times \partial\mathbf{E}^{\text{inc}}/\partial r$, and $\hat{\mathbf{r}} \times \mathbf{E}^{\text{inc}}$ can take on independent values at any particular point in space, each of the large square-bracketed quantities in (25b) must be zero, and thus equations (25) yield

$$\left[\frac{A_r(r)}{A_s(r)} - \frac{f(r)}{r\alpha_r(r)}\right] = 0, \qquad (26a)$$

$$\left[\frac{A_r(r)}{A_s(r)} - \frac{r\alpha_s(r)}{f(r)}\right] = 0, \qquad (26b)$$

$$\left[1 - \frac{\alpha_s(r)}{f'(r)}\right] = 0, \qquad (26c)$$

$$\left[\frac{1}{A_s(r)}\frac{\mathrm{d}A_s(r)}{\mathrm{d}r} + \frac{1}{r} - \frac{\alpha_s(r)}{f(r)}\right] = 0. \qquad (26d)$$

(The one exception to the possibility of the components of the field in (25) taking on independent values is if $\nabla \times \mathbf{E}^{\text{inc}}(\mathbf{r}) = 0$ (for $\mathbf{r}$ outside the source region), that is, if $\omega = 0$, in which case cloaking cannot be achieved by this formulation or by that in [1]. Cloaking of static fields is considered separately in section 5.) These four equations, along with the conditions in (17b), (18b) and (20b) are easily solved for $\alpha_s(r)$, $\alpha_r(r)$, $A_s(r)$, and $A_r(r)$ in terms of $f(r)$ to give (for $\mathbf{r} \in V$)

$$\alpha_s(r) = f'(r), \tag{27a}$$

$$\alpha_r(r) = \frac{1}{f'(r)} \left[\frac{f(r)}{r}\right]^2, \tag{27b}$$

$$A_s(r) = \frac{f(r)}{r}, \tag{27c}$$

$$A_r(r) = f'(r) \tag{27d}$$

with, in accordance with (17b), (18b) and (20b)

$$f(b) = b, \tag{28a}$$

$$f(a) = 0. \tag{28b}$$

Explicit expressions for the permittivity and permeability dyadics and for the fields are obtained by inserting (27) into (19), (21) and (20a); specifically

$$\bar{\boldsymbol{\alpha}} = \frac{\bar{\boldsymbol{\epsilon}}}{\epsilon_0} = \frac{\bar{\boldsymbol{\mu}}}{\mu_0} = \frac{1}{f'(r)} \left[\frac{f(r)}{r}\right]^2 \hat{\mathbf{r}}\hat{\mathbf{r}} + f'(r) \left(\hat{\boldsymbol{\theta}}\hat{\boldsymbol{\theta}} + \hat{\boldsymbol{\phi}}\hat{\boldsymbol{\phi}}\right), \tag{29a}$$

$$\mathbf{E}(\mathbf{r}) = \left[f'(r) - \frac{f(r)}{r}\right] E_r^{\text{inc}}[f(r), \theta, \phi]\hat{\mathbf{r}} + \frac{f(r)}{r} \mathbf{E}^{\text{inc}}[f(r), \theta, \phi], \tag{29b}$$

$$\mathbf{H}(\mathbf{r}) = \left[f'(r) - \frac{f(r)}{r}\right] H_r^{\text{inc}}[f(r), \theta, \phi]\hat{\mathbf{r}} + \frac{f(r)}{r} \mathbf{H}^{\text{inc}}[f(r), \theta, \phi], \tag{29c}$$

$$\frac{\mathbf{D}(\mathbf{r})}{\epsilon_0} = \bar{\boldsymbol{\alpha}}(\mathbf{r}) \cdot \mathbf{E}(\mathbf{r}) = \left[\left[\frac{f(r)}{r}\right]^2 - \frac{f(r)f'(r)}{r}\right] E_r^{\text{inc}}[f(r), \theta, \phi]\hat{\mathbf{r}}$$
$$+ \frac{f(r)f'(r)}{r} \mathbf{E}^{\text{inc}}[f(r), \theta, \phi], \tag{29d}$$

$$\frac{\mathbf{B}(\mathbf{r})}{\mu_0} = \bar{\boldsymbol{\alpha}}(\mathbf{r}) \cdot \mathbf{H}(\mathbf{r}) = \left[\left[\frac{f(r)}{r}\right]^2 - \frac{f(r)f'(r)}{r}\right] H_r^{\text{inc}}[f(r), \theta, \phi]\hat{\mathbf{r}}$$
$$+ \frac{f(r)f'(r)}{r} \mathbf{H}^{\text{inc}}[f(r), \theta, \phi]. \tag{29e}$$

We first note from (27) and (29) that without any distortion in the radial coordinate, $f(r) = r$ and the cloak reduces to free space. Secondly, the equations in (27) are independent of frequency if $f(r)$ is chosen independent of frequency and thus the corresponding relative permittivity and permeability ($\bar{\alpha}$) would be independent of frequency—a result that would violate causality-energy conditions; see section 6 below. Thirdly, Maxwell's equations for lossless media, along with $f(b) = b$, imply that the tangential components of the $\mathbf{E}(\mathbf{r})$ and $\mathbf{H}(\mathbf{r})$ fields are continuous across $r = b$ because $A_s(b) = 1$, and that the normal components of the $\mathbf{D}(\mathbf{r})$ and $\mathbf{B}(\mathbf{r})$ fields are continuous across $r = b$ because $\alpha_r(b) A_r(b) = 1$. Fourthly, the normal components of the $\mathbf{D}(\mathbf{r})$ and $\mathbf{B}(\mathbf{r})$ fields are zero at $r = a^+$ because $\alpha_r(a) A_r(a) = [f(a)/a]^2 = 0$, and the tangential components of the $\mathbf{E}(\mathbf{r})$ and $\mathbf{H}(\mathbf{r})$ fields are zero at $r = a^+$ because $A_s(a) = f(a)/a = 0$. Moreover, these boundary values are compatible with Maxwell's equations and the tangential $\mathbf{E}(\mathbf{r})$ and $\mathbf{H}(\mathbf{r})$ fields being continuous and thus zero across $r = a$ (no delta functions in the polarization at $r = a$; see section 2.1. With the tangential $\mathbf{E}(\mathbf{r})$ and $\mathbf{H}(\mathbf{r})$ fields zero just inside the surface of the free-space spherical cavity, Maxwell's equations demand that the fields everywhere inside the spherical cavity are zero. Lastly, observe from (29b), (29c) and (28b) that $(\mathbf{E} \times \mathbf{H}) \cdot \hat{\mathbf{r}} = (\mathbf{E} \times \mathbf{H}^*) \cdot \hat{\mathbf{r}} = 0$ at $r = a^+$. With $\mathbf{f}$, $\bar{\alpha}$ and $\bar{\mathbf{A}}$ independent of frequency, this implies from Poynting's theorem that time-domain incident fields are cloaked under the initial condition of zero fields in the cloak before the incident fields impinge upon the cloak; see section 2.1.

The equations in (27) with (28) determine an annular spherical cloak of inner radius $a$ and outer radius $b$ if the spherical region for $r < a$ is free space. And, of course, it is assumed that the infinite region $r > b$ outside the cloak is free space (except for the sources of the incident fields). For example, if in the region $a \leqslant r \leqslant b$

$$f(r) = \frac{b(r-a)^p}{(b-a)^p}, \quad p > 0, \tag{30a}$$

then

$$\alpha_s(r) = \frac{bp(r-a)^{p-1}}{(b-a)^p}, \tag{30b}$$

$$\alpha_r(r) = \frac{b(r-a)^{p+1}}{p(b-a)^p r^2}, \tag{30c}$$

$$A_s(r) = \frac{b(r-a)^p}{(b-a)^p r}, \tag{30d}$$

$$A_r(r) = \frac{bp(r-a)^{p-1}}{(b-a)^p}. \tag{30e}$$

For $p = 1$ these equations in (30) yield the spherical cloak of Pendry *et al* [1]. For $p > 1$, $A_r(a) = 0$ and the normal components of $\mathbf{E}(\mathbf{r})$ and $\mathbf{H}(\mathbf{r})$, that is, $E_r(\mathbf{r})$ and $H_r(\mathbf{r})$, are zero across $r = a$. (For $p = 1$, $E_r(\mathbf{r})$ and $H_r(\mathbf{r})$ are discontinuous across $r = a$.)

## 3.1. Spherical concentrators

If the boundary condition in (28b) is omitted, then the fields for $r < a$ are not generally zero and any continuous, piecewise continuously differentiable function $f(r)$, $0 \leqslant r \leqslant b$, in (27) satisfying (28a) will give zero scattered fields. These nonscattering spheres of radius $b$ differ from those in [14]–[19] in that the nonscattering is perfect and not restricted to bodies small enough that their electrical sizes lie in the dipolar or the low-order multipolar regime. (Although the cloaks of Kerker *et al* [14, 15] and Alu and Engheta [16]–[19] are generally smaller than a wavelength or two across, they have the advantage of not requiring anisotropic material.)

An interesting class of nonscattering spheres are defined by functions $f(r)$, $0 \leqslant r \leqslant b$ that concentrate and magnify the incident fields near the center of the scatterer. One particular coordinate function that concentrates and magnifies the incident fields inside a radius $a$ of the sphere (with outside radius $b$) is given by

$$f(r) = \begin{cases} Mr, & 0 \leqslant r \leqslant a, \\ \dfrac{(b-Ma)r + ab(M-1)}{b-a}, & a \leqslant r \leqslant b. \end{cases} \quad (31)$$

Note that $f(r)$ in (31) is continuous across $r = a$ so that $f'(r)$ does not produce any delta functions in the parameters of (27) when $f(r)$ is inserted to obtain

$$\alpha_s(r) = \begin{cases} M, & 0 \leqslant r < a, \\ \dfrac{b-Ma}{b-a}, & a < r \leqslant b, \end{cases} \quad (32a)$$

$$\alpha_r(r) = \begin{cases} M, & 0 \leqslant r < a, \\ \dfrac{[(b-Ma)r + ab(M-1)]^2}{(b-a)(b-Ma)r^2}, & a < r \leqslant b, \end{cases} \quad (32b)$$

$$A_s(r) = \begin{cases} M, & 0 \leqslant r \leqslant a, \\ \dfrac{(b-Ma)r + ab(M-1)}{(b-a)r}, & a \leqslant r \leqslant b, \end{cases} \quad (32c)$$

$$A_r(r) = \begin{cases} M, & 0 \leqslant r < a, \\ \dfrac{b-Ma}{b-a}, & a < r \leqslant b. \end{cases} \quad (32d)$$

Equations (32c) and (32d) show that $A_s = A_r = M$ for $0 \leqslant r < a$ and thus the incident fields for $r < a$ are magnified by a factor of $M$; that is,

$$\mathbf{E}(\mathbf{r}) = M\mathbf{E}^{\mathrm{inc}}(Mr, \theta, \phi), \quad 0 \leqslant r < a, \quad (33a)$$

$$\mathbf{H}(\mathbf{r}) = M\mathbf{H}^{\mathrm{inc}}(Mr, \theta, \phi), \quad 0 \leqslant r < a, \quad (33b)$$

while maintaining zero scattered fields for $r > b$. Equations (32a) and (32b) show that the relative permittivity and permeability are both equal to the same constant $M$ for $0 \leqslant r < a$; that is

$$\bar{\boldsymbol{\alpha}} = \dfrac{\bar{\boldsymbol{\epsilon}}}{\epsilon_0} = \dfrac{\bar{\boldsymbol{\mu}}}{\mu_0} = M\bar{\mathbf{I}}, \quad 0 \leqslant r < a. \quad (34)$$

## 4. Circular cylindrical cloaks

In this section, we consider a cloak consisting of an infinitely long circular cylindrical annulus of anisotropic material with inner radius $a$ and outer radius $b$. Such a cloak is conveniently described by the cylindrical coordinates $(u, v, w) = (\rho, \phi, z)$. In order for the circular cylindrical annulus to behave as a perfectly nonscattering cloak for all possible directions and values of the incident fields, which are assumed to have no variation in the $z$-direction, symmetry demands that

$$[\mathbf{f(r)}] = [f(\rho), g = \phi, h = z] \tag{35a}$$

with boundary condition (according to (16a))

$$f(b) = b \tag{35b}$$

and

$$\bar{\mathbf{A}}(\mathbf{r}) = A_\rho(\rho)\hat{\boldsymbol{\rho}}\hat{\boldsymbol{\rho}} + A_\phi(\rho)\hat{\boldsymbol{\phi}}\hat{\boldsymbol{\phi}} + A_z(\rho)\hat{\mathbf{z}}\hat{\mathbf{z}} \tag{36a}$$

with boundary conditions (according to (16b))

$$A_\phi(b) = A_z(b) = 1. \tag{36b}$$

Furthermore

$$\bar{\boldsymbol{\alpha}}(\mathbf{r}) = \alpha_\rho(\rho)\hat{\boldsymbol{\rho}}\hat{\boldsymbol{\rho}} + \alpha_\phi(\rho)\hat{\boldsymbol{\phi}}\hat{\boldsymbol{\phi}} + \alpha_z(\rho)\hat{\mathbf{z}}\hat{\mathbf{z}}, \tag{37}$$

so that

$$\bar{\boldsymbol{\alpha}}(\mathbf{r}) \cdot \bar{\mathbf{A}}(\mathbf{r}) = \alpha_\rho(\rho)A_\rho(\rho)\hat{\boldsymbol{\rho}}\hat{\boldsymbol{\rho}} + \alpha_\phi(\rho)A_\phi(\rho)\hat{\boldsymbol{\phi}}\hat{\boldsymbol{\phi}} + \alpha_z(\rho)A_z(\rho)\hat{\mathbf{z}}\hat{\mathbf{z}} \tag{38a}$$

with boundary condition (according to (16c))

$$\alpha_\rho(a)A_\rho(a) = 0 \tag{38b}$$

and

$$\bar{\mathbf{A}}(\mathbf{r}) \cdot \mathbf{E}^{\mathrm{inc}}[\mathbf{f(r)}] = A_\rho(\rho)E_\rho^{\mathrm{inc}}[f(\rho), \phi, z]\hat{\boldsymbol{\rho}} + A_\phi(\rho)E_\phi^{\mathrm{inc}}[f(\rho), \phi, z]\hat{\boldsymbol{\phi}} \\ + A_z(\rho)E_z^{\mathrm{inc}}[f(\rho), \phi, z]\hat{\mathbf{z}}. \tag{39}$$

Although the $z$-independent cylindrical fields uncouple into $E$ waves ($H_z = 0$) and $H$ waves ($E_z = 0$), the derivation remains simpler and more general if $E$- and $H$-wave fields are included together.

The curl of the incident field in cylindrical coordinates is given by

$$\nabla \times \mathbf{E}^{\mathrm{inc}}(\mathbf{r}) = \frac{1}{\rho}\frac{\partial E_z^{\mathrm{inc}}}{\partial \phi}\hat{\boldsymbol{\rho}} - \frac{\partial E_z^{\mathrm{inc}}}{\partial \rho}\hat{\boldsymbol{\phi}} + \left(\frac{\partial E_\phi^{\mathrm{inc}}}{\partial \rho} + \frac{E_\phi^{\mathrm{inc}}}{\rho} - \frac{1}{\rho}\frac{\partial E_\rho^{\mathrm{inc}}}{\partial \phi}\right)\hat{\mathbf{z}} \tag{40a}$$

and thus

$$\left[\nabla \times \mathbf{E}^{\mathrm{inc}}(\mathbf{r})\right]_{\mathbf{r} \Rightarrow \mathbf{f(r)}} = \frac{1}{f(\rho)}\frac{\partial E_z^{\mathrm{inc}}[f(\rho), \phi, z]}{\partial \phi}\hat{\boldsymbol{\rho}} - \frac{1}{f'(\rho)}\frac{\partial E_z^{\mathrm{inc}}[f(\rho), \phi, z]}{\partial \rho}\hat{\boldsymbol{\phi}} \\ + \left(\frac{1}{f'(\rho)}\frac{\partial E_\phi^{\mathrm{inc}}[f(\rho), \phi, z]}{\partial \rho} + \frac{E_\phi^{\mathrm{inc}}[f(\rho), \phi, z]}{f(\rho)} - \frac{1}{f(\rho)}\frac{\partial E_\rho^{\mathrm{inc}}[f(\rho), \phi, z]}{\partial \phi}\right)\hat{\mathbf{z}}, \tag{40b}$$

where $f'(\rho) = df(\rho)/d\rho$. From (39) and (40a) we can write

$$\nabla \times \left[\bar{\mathbf{A}}(\mathbf{r}) \cdot \mathbf{E}^{\text{inc}}[\mathbf{f}(\mathbf{r})]\right] = \frac{A_z(\rho)}{\rho} \frac{\partial E_z^{\text{inc}}[f(\rho), \phi, z]}{\partial \phi} \hat{\boldsymbol{\rho}} - \frac{\partial \left(A_z(\rho) E_z^{\text{inc}}[f(\rho), \phi, z]\right)}{\partial \rho} \hat{\boldsymbol{\phi}}$$
$$+ \left( \frac{\partial \left(A_\phi(\rho) E_\phi^{\text{inc}}[f(\rho), \phi, z]\right)}{\partial \rho} + \frac{A_\phi(\rho) E_\phi^{\text{inc}}[f(\rho), \phi, z]}{\rho} \right.$$
$$\left. - \frac{A_\rho(\rho)}{\rho} \frac{\partial E_\rho^{\text{inc}}[f(\rho), \phi, z]}{\partial \phi} \right) \hat{\mathbf{z}}. \tag{41}$$

Inserting (38a), (40b) and (41) into (15) and equating the $\hat{\boldsymbol{\rho}}$-, $\hat{\boldsymbol{\phi}}$- and $\hat{\mathbf{z}}$-components gives

$$\left[ \frac{A_z(\rho)}{\rho} - \frac{\alpha_\rho(\rho) A_\rho(\rho)}{f(\rho)} \right] \frac{\partial E_z^{\text{inc}}[f(\rho), \phi, z]}{\partial \phi} = 0, \tag{42a}$$

$$\left[ A_z(\rho) - \frac{\alpha_\phi(\rho) A_\phi(\rho)}{f'(\rho)} \right] \frac{\partial E_z^{\text{inc}}[f(\rho), \phi, z]}{\partial \rho} + A_z'(\rho) E_z^{\text{inc}}[f(\rho), \phi, z] = 0, \tag{42b}$$

$$\left[ A_\phi(\rho) - \frac{\alpha_z(\rho) A_z(\rho)}{f'(\rho)} \right] \frac{\partial E_\phi^{\text{inc}}[f(\rho), \phi, z]}{\partial \rho}$$
$$- \left[ \frac{A_\rho(\rho)}{\rho} - \frac{\alpha_z(\rho) A_z(\rho)}{f(\rho)} \right] \frac{\partial E_\rho^{\text{inc}}[f(\rho), \phi, z]}{\partial \phi}$$
$$+ \left[ \frac{dA_\phi(\rho)}{d\rho} + \frac{A_\phi(\rho)}{\rho} - \frac{\alpha_z(\rho) A_z(\rho)}{f(\rho)} \right] E_\phi^{\text{inc}}[f(\rho), \phi, z] = 0. \tag{42c}$$

Since the incident fields $\partial E_z^{\text{inc}}/\partial \rho$ and $E_z^{\text{inc}}$ can take on independent values at any particular point in space, and the incident fields $\partial E_\phi^{\text{inc}}/\partial \rho$, $\partial E_\rho^{\text{inc}}/\partial \phi$, and $E_\phi^{\text{inc}}$ can take on independent values at any particular point in space, each of the coefficients of these incident fields in (42b) and (42c) must be zero. Thus equations (42) yield

$$\left[ \frac{1}{\rho} - \frac{\alpha_\rho(\rho) A_\rho(\rho)}{f(\rho) A_z(\rho)} \right] = 0, \tag{43a}$$

$$\left[ 1 - \frac{\alpha_\phi(\rho) A_\phi(\rho)}{f'(\rho) A_z(\rho)} \right] = 0, \tag{43b}$$

$$A_z'(\rho) = 0, \tag{43c}$$

$$\left[ \frac{A_\phi(\rho)}{A_z(\rho)} - \frac{\alpha_z(\rho)}{f'(\rho)} \right] = 0, \tag{43d}$$

$$\left[ \frac{A_\rho(\rho)}{\rho A_z(\rho)} - \frac{\alpha_z(\rho)}{f(\rho)} \right] = 0, \tag{43e}$$

$$\left[ \frac{1}{A_z(\rho)} \left( \frac{dA_\phi(\rho)}{d\rho} + \frac{A_\phi(\rho)}{\rho} \right) - \frac{\alpha_z(\rho)}{f(\rho)} \right] = 0. \tag{43f}$$

$$A_z = 1 \tag{44a}$$

and thus the remaining five equations, along with the conditions in (35b), (36b) and (38b) are easily solved for $\alpha_z(\rho)$, $\alpha_\phi(\rho)$, $\alpha_\rho(r)$, $A_\phi(\rho)$ and $A_\rho(\rho)$ in terms of $f(\rho)$ to give (for $\mathbf{r} \in V$)

$$\alpha_z(\rho) = \frac{f(\rho) f'(\rho)}{\rho}, \tag{44b}$$

$$\alpha_\phi(\rho) = \frac{\rho f'(\rho)}{f(\rho)}, \tag{44c}$$

$$\alpha_\rho(\rho) = \frac{f(\rho)}{\rho f'(\rho)}, \tag{44d}$$

$$A_\phi(\rho) = \frac{f(\rho)}{\rho}, \tag{44e}$$

$$A_\rho(\rho) = f'(\rho) \tag{44f}$$

with, in accordance with (35b), (36b) and (38b)

$$f(b) = b, \tag{45a}$$

$$f(a) = 0. \tag{45b}$$

Explicit expressions for the permittivity and permeability dyadics and for the fields are obtained by inserting (44) into (37), (39) and (38a); specifically

$$\bar{\boldsymbol{\alpha}} = \frac{\bar{\boldsymbol{\epsilon}}}{\epsilon_0} = \frac{\bar{\boldsymbol{\mu}}}{\mu_0} = \frac{f(\rho)}{\rho f'(\rho)} \hat{\boldsymbol{\rho}} \hat{\boldsymbol{\rho}} + \frac{\rho f'(\rho)}{f(\rho)} \hat{\boldsymbol{\phi}} \hat{\boldsymbol{\phi}} + \frac{f(\rho) f'(\rho)}{\rho} \hat{\mathbf{z}} \hat{\mathbf{z}}, \tag{46a}$$

$$\mathbf{E}(\mathbf{r}) = f'(\rho) E_\rho^{\text{inc}}[f(\rho), \phi, z] \hat{\boldsymbol{\rho}} + \frac{f(\rho)}{\rho} E_\phi^{\text{inc}}[f(\rho), \phi, z] \hat{\boldsymbol{\phi}} + E_z^{\text{inc}}[f(\rho), \phi, z] \hat{\mathbf{z}}, \tag{46b}$$

$$\mathbf{H}(\mathbf{r}) = f'(\rho) H_\rho^{\text{inc}}[f(\rho), \phi, z] \hat{\boldsymbol{\rho}} + \frac{f(\rho)}{\rho} H_\phi^{\text{inc}}[f(\rho), \phi, z] \hat{\boldsymbol{\phi}} + H_z^{\text{inc}}[f(\rho), \phi, z] \hat{\mathbf{z}}, \tag{46c}$$

$$\frac{\mathbf{D}(\mathbf{r})}{\epsilon_0} = \bar{\boldsymbol{\alpha}}(\mathbf{r}) \cdot \mathbf{E}(\mathbf{r}) = \frac{f(\rho)}{\rho} E_\rho^{\text{inc}}[f(\rho), \phi, z] \hat{\boldsymbol{\rho}} + f'(\rho) E_\phi^{\text{inc}}[f(\rho), \phi, z] \hat{\boldsymbol{\phi}}$$
$$+ \frac{f(\rho) f'(\rho)}{\rho} E_z^{\text{inc}}[f(\rho), \phi, z] \hat{\mathbf{z}}, \tag{46d}$$

$$\frac{\mathbf{B}(\mathbf{r})}{\mu_0} = \bar{\boldsymbol{\alpha}}(\mathbf{r}) \cdot \mathbf{H}(\mathbf{r}) = \frac{f(\rho)}{\rho} H_\rho^{\text{inc}}[f(\rho), \phi, z] \hat{\boldsymbol{\rho}} + f'(\rho) H_\phi^{\text{inc}}[f(\rho), \phi, z] \hat{\boldsymbol{\phi}}$$
$$+ \frac{f(\rho) f'(\rho)}{\rho} H_z^{\text{inc}}[f(\rho), \phi, z] \hat{\mathbf{z}}. \tag{46e}$$

We first note from (44) and (46) that without any distortion in the radial coordinate, $f(\rho) = \rho$ and the cloak reduces to free space. In general, however, as the $\rho$ components of the permittivity–permeability dyadics approach a value of zero as $\rho \to a$ (similar to the $r$-components for the spherical cloak), the $\phi$-components always approach an infinite value as $\rho \to a$ (a requirement that may make cylindrical cloaking more difficult to approximate than spherical cloaking). Secondly, the equations in (44) are independent of frequency if $f(\rho)$ is chosen independent of frequency and thus the corresponding relative permittivity and permeability ($\bar{\boldsymbol{\alpha}}$) would be independent of frequency—a result that would violate causality-energy conditions; see section 6 below. Thirdly, Maxwell's equations for lossless media, along with $f(b) = b$, imply that the tangential components of the $\mathbf{E}(\mathbf{r})$ and $\mathbf{H}(\mathbf{r})$ fields are continuous across $\rho = b$ because $A_z(b) = A_\phi(b) = 1$, and that the normal components of the $\mathbf{D}(\mathbf{r})$ and $\mathbf{B}(\mathbf{r})$ fields $[D_\rho(\mathbf{r}), B_\rho(\mathbf{r})]$ are continuous across $\rho = b$ because $\alpha_\rho(b) A_\rho(b) = 1$. Fourthly, the normal components of the $\mathbf{D}(\mathbf{r})$ and $\mathbf{B}(\mathbf{r})$ fields $[D_\rho(\mathbf{r}), B_\rho(\mathbf{r})]$ are zero at $\rho = a^+$ because $\alpha_\rho(a) A_\rho(a) = f(a)/a = 0$, and also $E_\phi(\mathbf{r})$ and $H_\phi(\mathbf{r})$ are zero at $\rho = a^+$ because $A_\phi(a) = f(a)/a = 0$. In addition, $E_z(a^+, \phi) = E_z^{\text{inc}}[f(a) = 0, \phi] = E_z^{\text{inc}}(\rho = 0)$, which is independent of $\phi$; that is $E_z(\rho = a^+)$ (and likewise $H_z(\rho = a^+)$) is independent of $\phi$. Moreover, these boundary values are compatible with Maxwell's equations and the $[E_\phi(\mathbf{r}), H_\phi(\mathbf{r})]$ and $[B_\rho(\mathbf{r}), D_\rho(\mathbf{r})]$ fields being continuous and thus zero across $\rho = a$ (no delta functions in polarization at $\rho = a$ giving rise to these field components); see section 2.1. With these boundary conditions, the $\rho$- and $\phi$-components of the $\mathbf{E}(\mathbf{r})$ and $\mathbf{H}(\mathbf{r})$ fields are zero just inside the surface of the free-space cylindrical cavity and a cylindrical mode expansion shows that the fields everywhere inside the free-space cylindrical cavity must be zero. Lastly, observe from (46b), (46c) and (45b) that $(\mathbf{E} \times \mathbf{H}) \cdot \hat{\boldsymbol{\rho}} = (\mathbf{E} \times \mathbf{H}^*) \cdot \hat{\boldsymbol{\rho}} = 0$ at $\rho = a^+$. With $\mathbf{f}$, $\bar{\boldsymbol{\alpha}}$ and $\bar{\mathbf{A}}$ independent of frequency, this implies from Poynting's theorem that time-domain incident fields are cloaked under the initial condition of zero fields in the cloak before the incident fields impinge upon the cloak; see section 2.1.

*Emphatically, however, $E_z(\mathbf{r})$ and $H_z(\mathbf{r})$ are not zero as $\rho \to a$ from inside V because $A_z(a) = 1$ and thus the solution compatible with zero fields inside the free-space cavity has z tangential components of the $\mathbf{E}(\mathbf{r})$ and $\mathbf{H}(\mathbf{r})$ fields that are discontinuous across $\rho = a$. In other words, unlike the spherical cloak, all the tangential components of the $\mathbf{E}(\mathbf{r})$ and $\mathbf{H}(\mathbf{r})$ fields of the cylindrical cloak are not continuous across the inner surface defined by $\rho = a$.* This discontinuity of $E_z(\mathbf{r})$ and $H_z(\mathbf{r})$ across the inner surface of the cylindrical cloak at $\rho = a$ implies from Maxwell's equations (1) that the curls of $\mathbf{E}(\mathbf{r})$ and $\mathbf{H}(\mathbf{r})$ contain delta functions at $\rho = a$. Since the permittivity and permeability of the cloak are lossless, there can be no electric or magnetic surface currents at $\rho = a$ and thus (1)–(2) show that these delta functions in the curls of (1) give rise to delta functions in the $B_\phi(\mathbf{r})$ and $D_\phi(\mathbf{r})$ fields at $\rho = a$ (that is, delta functions in the $\phi$ components of polarization densities at $\rho = a$). The delta functions in $\phi$ polarization are compatible with $\alpha_\phi = \rho f'(\rho)/f(\rho)$, the relative permittivity and permeability in the $\phi$ direction, because from the mean value theorem for continuous differentiable functions with $f(a) = 0$

$$\lim_{\rho \to a} \alpha_\phi(\rho) = \lim_{\rho \to a} \frac{\rho f'(\rho)}{f(\rho)} = \lim_{\rho \to a} \frac{\rho}{\rho - a} = \infty \,. \tag{47}$$

These delta functions at $\rho = a$ for the cylindrical cloak were first found by Greenleaf *et al* [6], who showed that they spread out to large but finite values of the $B_\phi(\mathbf{r})$ and $D_\phi(\mathbf{r})$ fields as $\rho$ gets close to $a$ from inside V if the permeability or permittivity of the cloak differs slightly from their perfect cloaking values.

The equations in (44) with (45) determine a circular cylinder annular cloak of inner radius $a$ and outer radius $b$ if the cylindrical region for $\rho < a$ is free space. And, of course, it is assumed that the infinite region $\rho > b$ outside the cloak is free space (except for the sources of the incident fields). For example, if in the region $a \leqslant \rho \leqslant b$

$$f(\rho) = \frac{b(\rho - a)^p}{(b - a)^p}, \quad p > 0, \tag{48a}$$

then

$$\alpha_z(\rho) = \frac{b^2 p (\rho - a)^{2p-1}}{(b - a)^{2p} \rho}, \tag{48b}$$

$$\alpha_\phi(\rho) = \frac{p\rho}{\rho - a}, \tag{48c}$$

$$\alpha_\rho(\rho) = \frac{\rho - a}{p\rho}, \tag{48d}$$

$$A_z(\rho) = 1, \tag{48e}$$

$$A_\phi(\rho) = \frac{b(\rho - a)^p}{(b - a)^p \rho}, \tag{48f}$$

$$A_\rho(\rho) = \frac{bp(\rho - a)^{p-1}}{(b - a)^p}. \tag{48g}$$

For $p = 1$ these equations in (48) yield the cylindrical cloak of Cummer *et al* [11]. For $p > 1$, $A_\rho(a) = 0$ and the normal components of $\mathbf{E}(\mathbf{r})$ and $\mathbf{H}(\mathbf{r})$, that is, $E_\rho(\mathbf{r})$ and $H_\rho(\mathbf{r})$, are zero across $\rho = a$. (For $p = 1$, $E_\rho(\mathbf{r})$ and $H_\rho(\mathbf{r})$ are discontinuous across $\rho = a$.)

Finally, we note that if the condition $f(b) = b$ is omitted, the circular cylindrical annulus will scatter but still produce zero fields in the region $\rho < a$. For $H$-wave incident fields, we can choose $f(\rho) = \sqrt{\rho^2 - a^2}$ to yield one such scattering circular cylindrical annulus with

$$\frac{\bar{\mu}}{\mu_0} = \bar{\mathbf{I}} \tag{49a}$$

and

$$\frac{\bar{\epsilon}}{\epsilon_0} = \frac{\rho^2 - a^2}{\rho^2} \hat{\boldsymbol{\rho}}\hat{\boldsymbol{\rho}} + \frac{\rho^2}{\rho^2 - a^2} \hat{\boldsymbol{\phi}}\hat{\boldsymbol{\phi}} + \hat{\mathbf{z}}\hat{\mathbf{z}}. \tag{49b}$$

That is, the annulus is nonmagnetic with only the permittivity dyadic different from that of free space. However, a simple perfectly conducting shell is also a scatterer with zero interior cavity fields. A nonmagnetic circular cylinder annulus that behaves as an approximate cloak (nonzero scattered and nonzero interior cavity fields) at optical frequencies has recently been designed [31] and experimentally realized [32].

*4.1. Circular cylindrical concentrators*

If the boundary condition in (45b) is omitted, then the fields for $\rho < a$ are not generally zero and any continuous, piecewise continuously differentiable function $f(\rho)$, $0 \leqslant \rho \leqslant b$, in (44) satisfying (45a) will give zero scattered fields. An interesting class of nonscattering circular cylinders are defined by functions $f(\rho)$, $0 \leqslant \rho \leqslant b$ that concentrate and magnify the incident fields near the center of the scatterer [33]. One particular coordinate function that concentrates and magnifies the incident fields inside a radius $a$ of the circular cylinder (with outside radius $b$) is given by

$$f(\rho) = \begin{cases} M\rho, & 0 \leqslant \rho \leqslant a, \\ \dfrac{(b-Ma)\rho + ab(M-1)}{b-a}, & a \leqslant \rho \leqslant b. \end{cases} \quad (50)$$

Note that $f(\rho)$ in (50) is continuous across $\rho = a$ so that $f'(\rho)$ does not produce any delta functions in the parameters of (44) when $f(\rho)$ is inserted to obtain

$$\alpha_z(\rho) = \begin{cases} M^2, & 0 \leqslant \rho \leqslant a, \\ \dfrac{[(b-Ma)\rho + ab(M-1)](b-Ma)}{(b-a)^2 \rho}, & a \leqslant \rho \leqslant b, \end{cases} \quad (51a)$$

$$\alpha_\phi(\rho) = \begin{cases} 1, & 0 \leqslant \rho < a, \\ \dfrac{(b-Ma)\rho}{(b-Ma)\rho + ab(M-1)}, & a < \rho \leqslant b, \end{cases} \quad (51b)$$

$$\alpha_\rho(\rho) = \begin{cases} 1, & 0 \leqslant \rho < a, \\ \dfrac{(b-Ma)\rho + ab(M-1)}{(b-Ma)\rho}, & a < \rho \leqslant b, \end{cases} \quad (51c)$$

$$A_z = 1, \quad 0 \leqslant \rho \leqslant b, \quad (51d)$$

$$A_\phi(\rho) = \begin{cases} M, & 0 \leqslant \rho \leqslant a, \\ \dfrac{(b-Ma)\rho + ab(M-1)}{(b-a)\rho}, & a \leqslant \rho \leqslant b, \end{cases} \quad (51e)$$

$$A_\rho(\rho) = \begin{cases} M, & 0 \leqslant \rho < a, \\ \dfrac{b-Ma}{b-a}, & a < \rho \leqslant b. \end{cases} \quad (51f)$$

Equations (51e) and (51f) show that $A_\phi = A_\rho = M$ for $0 \leqslant \rho < a$ and thus the associated components of the incident field for $\rho < a$ are magnified by a factor of $M$; that is,

$$\mathbf{E}(\mathbf{r}) = M\left[E_\rho^{\text{inc}}(M\rho, \phi, z)\hat{\boldsymbol{\rho}} + E_\phi^{\text{inc}}(M\rho, \phi, z)\hat{\boldsymbol{\phi}}\right] + E_z^{\text{inc}}(M\rho, \phi, z)\hat{\mathbf{z}}, \quad 0 \leqslant \rho < a, \quad (52a)$$

$$\mathbf{H}(\mathbf{r}) = M\left[H_\rho^{\text{inc}}(M\rho, \phi, z)\hat{\boldsymbol{\rho}} + H_\phi^{\text{inc}}(M\rho, \phi, z)\hat{\boldsymbol{\phi}}\right] + H_z^{\text{inc}}(M\rho, \phi, z)\hat{\mathbf{z}}, \quad 0 \leqslant \rho < a, \quad (52b)$$

while maintaining zero scattered fields for $\rho > b$. Equations (51a)–(51c) show that the relative permittivity and permeability dyadics in the region $0 \leqslant \rho < a$ are given by

$$\bar{\boldsymbol{\alpha}} = \frac{\bar{\boldsymbol{\epsilon}}}{\epsilon_0} = \frac{\bar{\boldsymbol{\mu}}}{\mu_0} = \hat{\boldsymbol{\rho}}\hat{\boldsymbol{\rho}} + \hat{\boldsymbol{\phi}}\hat{\boldsymbol{\phi}} + M^2 \hat{\mathbf{z}}\hat{\mathbf{z}}. \quad (53)$$

A similar circular cylindrical concentrator was first derived by Rahm *et al* [33].

## 5. Cloaking of static fields

For static fields, $\omega = 0$ and the equations in (1) reduce, outside the sources of the incident fields, to simply

$$\nabla \times \mathbf{E}(\mathbf{r}) = 0, \tag{54a}$$

$$\nabla \times \mathbf{H}(\mathbf{r}) = 0. \tag{54b}$$

The divergence equations

$$\nabla \cdot \mathbf{D}(\mathbf{r}) = 0, \tag{55a}$$

$$\nabla \cdot \mathbf{B}(\mathbf{r}) = 0 \tag{55b}$$

must augment (54a) and (54b), respectively. Thus, the cloaking equations derived in previous sections for $\omega > 0$ do not apply and, in particular, the relative permittivity–permeability dyadic $\bar{\boldsymbol{\alpha}}(\mathbf{r})$ satisfying (15) and (16) does not cloak static fields. Nonetheless, we can use the same straightforward, boundary value method to derive a permittivity–permeability dyadic for cloaking static fields. (The similarity between the electrostatic equations ((54a) and (55a)), and the magnetostatic equations ((54b) and (55b)), demands that a relative permittivity dyadic that cloaks electrostatic fields is identical to a relative permittivity dyadic that cloaks magnetostatic fields.) However, we shall find in sections 5.1 and 5.2 that some of the elements of the relative permittivity dyadic for spherical and cylindrical cloaks, unlike the elements of the relative permeability dyadic, violate causality-energy conditions as $\omega \to 0$ because magnetic current loops do not exist that can produce 'diaelectric' polarization in the way that electric current loops can produce diamagnetic polarization. Consequently, we shall use the magnetostatic differential equations, (54b) and (55b), in the following derivation for the cloaking of static fields. (It should be mentioned that we are using the term 'cloak' for static fields in the same way as throughout the rest of the paper, namely, to refer to a lossless annulus with free-space inside and out. Static 'cloaking' of electric and magnetic fields has been formulated for conductive (lossy) bodies in conductive background media [26, 27].)

Assume an anisotropic lossless cloaking material with the constitutive relationship given in (2b), rewritten in terms of a relative permeability dyadic $\bar{\boldsymbol{\alpha}}(\mathbf{r})$ as

$$\mathbf{B}(\mathbf{r}) = \mu_0 \bar{\boldsymbol{\alpha}}(\mathbf{r}) \cdot \mathbf{H}(\mathbf{r}), \tag{56}$$

so that

$$\mathbf{H}(\mathbf{r}) = \frac{1}{\mu_0} \bar{\boldsymbol{\alpha}}^{-1}(\mathbf{r}) \cdot \mathbf{B}(\mathbf{r}) \tag{57}$$

and (54b) becomes

$$\nabla \times \left[ \bar{\boldsymbol{\alpha}}^{-1}(\mathbf{r}) \cdot \mathbf{B}(\mathbf{r}) \right] = 0, \tag{58}$$

within the volume $V$ of the cloaking material. Also, assume that the magnetic induction inside the volume $V$ takes the form

$$\mathbf{B}(\mathbf{r}) = \bar{\mathbf{A}}(\mathbf{r}) \cdot \mathbf{B}^{\text{inc}}[\mathbf{f}(\mathbf{r})]. \tag{59}$$

We thus want to find $\bar{\boldsymbol{\alpha}}(\mathbf{r})$ and $\bar{\mathbf{A}}(\mathbf{r})$ in terms of $\mathbf{f}(\mathbf{r})$ that will satisfy (58) for all possible $\mathbf{B}^{\text{inc}}(\mathbf{r})$

$$\nabla \times \left\{ \bar{\boldsymbol{\alpha}}^{-1}(\mathbf{r}) \cdot \bar{\mathbf{A}}(\mathbf{r}) \cdot \mathbf{B}^{\text{inc}}[\mathbf{f}(\mathbf{r})] \right\} = 0, \quad \mathbf{r} \in V, \tag{60a}$$

along with (55b), that is

$$\nabla \cdot \{\bar{\mathbf{A}}(\mathbf{r}) \cdot \mathbf{B}^{\text{inc}}[\mathbf{f}(\mathbf{r})]\} = 0, \tag{60b}$$

where, of course

$$\nabla \times \mathbf{B}^{\text{inc}}(\mathbf{r}) = 0, \quad \mathbf{r} \in V, \tag{60c}$$

$$\nabla \cdot \mathbf{B}^{\text{inc}}(\mathbf{r}) = 0 \tag{60d}$$

and, as before, $\mathbf{f}(\mathbf{r})$ must obey the following boundary condition at the outer boundary of the cloak:

$$\mathbf{f}(\mathbf{r}) \stackrel{\mathbf{r} \to S_b^-}{=} \mathbf{r}. \tag{61a}$$

To ensure that there is no surface current on the outer boundary $S_b$ of the cloak, the tangential magnetic field $\mathbf{H}(\mathbf{r})$ is required to be continuous across $S_b$, that is

$$\hat{\mathbf{n}} \times [\bar{\boldsymbol{\alpha}}^{-1}(\mathbf{r}) \cdot \bar{\mathbf{A}}(\mathbf{r})] \stackrel{\mathbf{r} \to S_b^-}{=} \hat{\mathbf{n}} \times \bar{\mathbf{I}}. \tag{61b}$$

Since (60a) and (60b) are applied separately, continuity of the normal component of $\mathbf{B}(\mathbf{r})$ must also be imposed across the outer boundary of the cloak to ensure (through the uniqueness theorem for static fields) that the scattered magnetostatic fields will be zero; specifically

$$\hat{\mathbf{n}} \cdot \bar{\mathbf{A}}(\mathbf{r}) \stackrel{\mathbf{r} \to S_b^-}{=} \hat{\mathbf{n}} \cdot \bar{\mathbf{I}} = \hat{\mathbf{n}}. \tag{61c}$$

The boundary condition as $\mathbf{r} \to S_a^+$ must be compatible with zero fields inside $S_a$ and ensure that there are no surface currents on the inner boundary $S_a$. This can be achieved by requiring that the tangential magnetic field $\mathbf{H}(\mathbf{r})$ be zero as $\mathbf{r} \to S_a^+$, that is

$$\hat{\mathbf{n}} \times [\bar{\boldsymbol{\alpha}}^{-1}(\mathbf{r}) \cdot \bar{\mathbf{A}}(\mathbf{r})] \stackrel{\mathbf{r} \to S_a^+}{=} 0. \tag{61d}$$

## 5.1. Spherical magnetostatic cloak

For the spherical cloak, symmetry requires that

$$[\mathbf{f}(\mathbf{r})] = [f(\rho), g = \theta, h = \phi], \tag{62a}$$

$$\bar{\mathbf{A}}(\mathbf{r}) = A_r(r)\hat{\mathbf{r}}\hat{\mathbf{r}} + A_s(r)\left(\hat{\boldsymbol{\theta}}\hat{\mathbf{r}}h + \hat{\boldsymbol{\phi}}\hat{\boldsymbol{\phi}}\right) \tag{62b}$$

and

$$\bar{\boldsymbol{\alpha}}(\mathbf{r}) = \alpha_r(r)\hat{\mathbf{r}}\hat{\mathbf{r}} + \alpha_s(r)\left(\hat{\boldsymbol{\theta}}\hat{\boldsymbol{\theta}} + \hat{\boldsymbol{\phi}}\hat{\boldsymbol{\phi}}\right) \tag{62c}$$

with the boundary conditions in (61) giving

$$f(b) = b, \tag{63a}$$

$$\frac{A_s(b)}{\alpha_s(b)} = 1, \tag{63b}$$

$$A_r(b) = 1 \tag{63c}$$

and

$$\frac{A_s(a)}{\alpha_s(a)} = 0. \tag{63d}$$

Inserting the functions in (62) into the equations in (60) and proceeding as in section 3, we find

$$\alpha_s(r) = \frac{rf'(r)}{f(r)}, \tag{64a}$$

$$\alpha_r(r) = \frac{f(r)}{rf'(r)}, \tag{64b}$$

$$A_s(r) = f'(r), \tag{64c}$$

$$A_r(r) = \frac{f(r)}{r} \tag{64d}$$

with the boundary conditions

$$f(b) = b, \tag{65a}$$

$$f(a) = 0. \tag{65b}$$

The boundary condition in (65b) implies from (64a) and (64b) that the values of the tangential components of the permeability dyadic approach infinity as $r \to a^+$, whereas the value of the radial component goes to zero as $r \to a^+$. However, all the components of $\mathbf{B}(\mathbf{r})$ and $\mathbf{H}(\mathbf{r})$ remain finite as $r \to a^+$ with the radial component of $\mathbf{B}(\mathbf{r})$ and the tangential components of $\mathbf{H}(\mathbf{r})$ going to zero as $r \to a^+$. The general expressions for $\mathbf{B}(\mathbf{r})$ and $\mathbf{H}(\mathbf{r})$ are

$$\mathbf{B}(\mathbf{r}) = \frac{f(r)}{r} B_r^{\text{inc}}[f(r), \theta, \phi]\hat{\mathbf{r}} - f'(r)\hat{\mathbf{r}} \times \hat{\mathbf{r}} \times \mathbf{B}^{\text{inc}}[f(r), \theta, \phi], \tag{66a}$$

$$\mu_0 \mathbf{H}(\mathbf{r}) = f'(r) B_r^{\text{inc}}[f(r), \theta, \phi]\hat{\mathbf{r}} - \frac{f(r)}{r}\hat{\mathbf{r}} \times \hat{\mathbf{r}} \times \mathbf{B}^{\text{inc}}[f(r), \theta, \phi]. \tag{66b}$$

As an example, if in the region $a \leqslant r \leqslant b$

$$f(r) = \frac{b(r-a)^p}{(b-a)^p}, \quad p > 0, \tag{67a}$$

then

$$\alpha_s(r) = \frac{pr}{r-a}, \tag{67b}$$

$$\alpha_r(r) = \frac{r-a}{pr}, \tag{67c}$$

$$A_s(r) = \frac{pb(r-a)^{p-1}}{(b-a)^p}, \tag{67d}$$

$$A_r(r) = \frac{b(r-a)^p}{r(b-a)^p}. \tag{67e}$$

## 5.2. Cylindrical magnetostatic cloak

For the cylindrical cloak, the vanishing curl of the incident magnetic field implies that $H_z^{\text{inc}}$ is a constant throughout space. Thus, we can assume that the constant is zero and consider only the $\rho$- and $\phi$-components of the magnetic field and magnetic induction. Symmetry then requires that

$$[\mathbf{f}(\mathbf{r})] = [f(\rho), g = \phi, h = z], \tag{68a}$$

$$\bar{\mathbf{A}}(\mathbf{r}) = A_\rho(\rho)\hat{\boldsymbol{\rho}}\hat{\boldsymbol{\rho}} + A_\phi(\rho)\hat{\boldsymbol{\phi}}\hat{\boldsymbol{\phi}} \tag{68b}$$

and

$$\bar{\boldsymbol{\alpha}}(\mathbf{r}) = \alpha_\rho(\rho)\hat{\boldsymbol{\rho}}\hat{\boldsymbol{\rho}} + \alpha_\phi(\rho)\hat{\boldsymbol{\phi}}\hat{\boldsymbol{\phi}} \tag{68c}$$

with the boundary conditions in (61) giving

$$f(b) = b, \tag{69a}$$

$$\frac{A_\phi(b)}{\alpha_\phi(b)} = 1, \tag{69b}$$

$$A_\rho(b) = 1, \tag{69c}$$

and

$$\frac{A_\phi(a)}{\alpha_\phi(a)} = 0. \tag{69d}$$

Inserting the functions in (68) into the equations in (60) and proceeding as in section 3, we find

$$\alpha_\phi(\rho) = \frac{\rho f'(\rho)}{f(\rho)}, \tag{70a}$$

$$\alpha_\rho(\rho) = \frac{f(\rho)}{\rho f'(\rho)}, \tag{70b}$$

$$A_\phi(\rho) = f'(\rho), \tag{70c}$$

$$A_\rho(\rho) = \frac{f(\rho)}{\rho} \tag{70d}$$

with the boundary conditions

$$f(b) = b, \tag{71a}$$

$$f(a) = 0. \tag{71b}$$

The boundary condition in (71b) implies from (70a) and (70b) that the values of the tangential component of the permeability dyadic approaches infinity as $\rho \to a^+$ while the value of the radial component goes to zero as $\rho \to a^+$. However, all the components of $\mathbf{B}(\mathbf{r})$ and $\mathbf{H}(\mathbf{r})$ remain finite as $\rho \to a^+$ with the radial component of $\mathbf{B}(\mathbf{r})$ and the tangential component of $\mathbf{H}(\mathbf{r})$ going to zero as $\rho \to a^+$. The general expressions for $\mathbf{B}(\mathbf{r})$ and $\mathbf{H}(\mathbf{r})$ are

$$\mathbf{B}(\mathbf{r}) = \frac{f(\rho)}{\rho} B_\rho^{\text{inc}}[f(\rho), \phi, z]\hat{\boldsymbol{\rho}} + f'(\rho) B_\phi^{\text{inc}}[f(\rho), \phi, z]\hat{\boldsymbol{\phi}}, \qquad (72a)$$

$$\mu_0 \mathbf{H}(\mathbf{r}) = f'(\rho) B_\rho^{\text{inc}}[f(\rho), \phi, z]\hat{\boldsymbol{\rho}} + \frac{f(\rho)}{\rho} B_\phi^{\text{inc}}[f(\rho), \phi, z]\hat{\boldsymbol{\phi}}. \qquad (72b)$$

As an example, if in the region $a \leqslant \rho \leqslant b$

$$f(\rho) = \frac{b(\rho - a)^p}{(b-a)^p}, \quad p > 0, \qquad (73a)$$

then

$$\alpha_\phi(\rho) = \frac{p\rho}{\rho - a}, \qquad (73b)$$

$$\alpha_\rho(\rho) = \frac{\rho - a}{p\rho}, \qquad (73c)$$

$$A_\phi(\rho) = \frac{pb(\rho - a)^{p-1}}{(b-a)^p}, \qquad (73d)$$

$$A_\rho(\rho) = \frac{b(\rho - a)^p}{\rho(b-a)^p}. \qquad (73e)$$

Note that the equations for the static-field cylindrical cloak, unlike those for the $\omega > 0$ cylindrical cloak, have the same form as the equations for the spherical cloak ($\rho$ simply replacing $r$). Also, comparing (70a) and (70b) with (44c) and (44d) shows that $\alpha_\phi(\rho)$ and $\alpha_\rho(\rho)$ for the static ($\omega = 0$) and time-harmonic ($\omega > 0$) cylindrical cloaks are identical.

## 6. Implications of causality-energy conditions

We shall show in this section that causality-energy conditions, which the diagonal elements of the relative permittivity–permeability dyadic must obey, imply that incident fields with a finite (nonzero) bandwidth cannot be perfectly cloaked. The causality-energy conditions for permittivity as $\omega \to 0$ are also used to show that electrostatic fields and low-frequency fields for $\omega > 0$, unlike magnetostatic fields, cannot be cloaked. This result is a consequence of the causality-energy conditions for permittivity being more restrictive than those for permeability because magnetic dipoles are caused by circulating electric currents rather than by the separation of magnetic charge, whereas electric dipoles are caused by the separation of electric charge rather than circulating magnetic current.

What is usually referred to as 'causality' conditions, we refer to as 'causality-energy' conditions because we find it difficult to justify their traditional derivation [34, section 84] from the Kramers–Kronig dispersion relations. This traditional derivation from the Kramers–Kronig relations depends on the macroscopic elements of the permittivity and permeability dyadics being well defined for all frequencies and yet all real materials (and metamaterials) fail to be describable by macroscopic constitutive relations if the wavelength of the fields is on the order of the distance between the microscopic dipoles. Moreover, above a certain frequency, induced circulating currents that give rise to diamagnetism can no longer be described by a spatially nondispersive permeability dyadic [34, sections 79, 82 and 103]. Therefore, we rely on the derivation from electromagnetic energy conservation [35]–[37] to obtain the following inequalities for the diagonal elements of the relative permittivity or permeability dyadic in a frequency window where there is no loss

$$\frac{\mathrm{d}(\omega\alpha_{ll})}{\mathrm{d}\omega} - 1 \geqslant \frac{\omega}{2}\frac{\mathrm{d}\alpha_{ll}}{\mathrm{d}\omega} \geqslant 0, \quad l = 1, 2, 3. \tag{74}$$

For the spherical cloak $\alpha_{ll} = (\alpha_r, \alpha_s, \alpha_s)$ and for the cylindrical cloak $\alpha_{ll} = (\alpha_\rho, \alpha_\phi, \alpha_z)$. From (27b) and (28b), or (44d) and (45b), we see that (74) must hold for the spherical cloak with $\alpha_r(a) = 0$ and for the cylindrical cloak with $\alpha_\rho(a) = 0$ at all frequencies. Thus, even if $f(r)$ and $f(\rho)$ are allowed to vary with frequency, it is impossible for all the inequalities in (74) to hold at any frequency. That is, an ideal spherical or cylindrical cloak (nonscattering and zero total interior cavity fields) over any nonzero bandwidth always violates causality and thus the cloaking of realistic (time-dependent) incident fields must be approximate.

For a generally shaped annular cloak, the energy in any finite length pulse (fields with nonzero bandwidth) must travel through the cloak in the same time that it would take without the cloak present since the cloak does not scatter the incident fields. Consequently, the energy in the pulse, which must bend around the inner cavity of the cloak, would have to travel faster than the free-space speed of light inside the cloak material and thus violate causality [29]. Again this implies that the cloaking of nonzero bandwidth fields must be approximate, though the approximation may be a very good one for sinusoidal pulse lengths significantly larger in space than the largest dimension of the cloak. Even in that case, nonetheless, the transient fields from the leading and trailing edges of this lengthy sinusoidal pulse will not be cloaked.

*6.1. Causality-energy conditions for diamagnetic materials (and metamaterials)*

The inequalities in (74) were derived from electromagnetic energy conservation based on the following electromagnetic power relation for polarized material [35]–[37]

$$P_{\mathrm{e}\ell}(\mathbf{r},t) = \frac{\partial \boldsymbol{\mathcal{D}}}{\partial t} \cdot \boldsymbol{\mathcal{E}} + \frac{\partial \boldsymbol{\mathcal{B}}}{\partial t} \cdot \boldsymbol{\mathcal{H}} - \frac{1}{2}\frac{\partial}{\partial t}\left(\epsilon_0|\boldsymbol{\mathcal{E}}|^2 + \mu_0|\boldsymbol{\mathcal{H}}|^2\right). \tag{75}$$

It was proven in [38, section 2.1.10] that this $P_{\mathrm{e}\ell}(\mathbf{r},t)$ is equal to the power per unit volume supplied to the polarization by the electromagnetic fields at the time $t$. The proof depended on showing that the fields supplied an internal power contribution to pre-existing (as in paramagnetic and ferro(i)magnetic materials) Amperian ('current loop') magnetic dipoles equal to $\partial(\mathbf{m}\cdot\mathbf{B}_0)/\partial t$, where $\mathbf{m}$ is the Amperian dipole moment and $\mathbf{B}_0$ is the local magnetic induction applied to the magnetic dipole; see [38, equation (2.162)]. If the Amperian magnetic dipoles do not exist before the local field is applied, that is, the applied field induces the Amperian magnetic dipole moments (as in diamagnetic materials or metamaterials) and does not just align pre-existing Amperian magnetic dipole moments (as in paramagnetic and ferro(i)magnetic materials), this contribution can be zero and thus the power relation that becomes relevant for obtaining inequalities for diamagnetic materials or metamaterials corresponding to (74) is given

by [38, equation (2.139)]

$$P_{e\ell}^0(\mathbf{r}, t) = \frac{\partial \mathcal{D}}{\partial t} \cdot \mathcal{E} + \frac{\partial \mathcal{B}}{\partial t} \cdot \mathcal{H} - \frac{1}{2} \frac{\partial}{\partial t} \left(\epsilon_0 |\mathcal{E}|^2 + |\mathcal{B}|^2/\mu_0\right). \quad (76)$$

Using (76) instead of (75) in the electromagnetic energy conservation derivation contained in [35]–[37], we obtain the same inequalities as in (74) for the diagonal elements of a lossless relative permittivity dyadic, but the following inequalities for diamagnetic materials or metamaterials:

$$\frac{d(\omega \alpha_{ll}^d)}{d\omega} - \left(\alpha_{ll}^d\right)^2 \geqslant \frac{\omega}{2} \frac{d\alpha_{ll}^d}{d\omega} \geqslant 0, \quad l = 1, 2, 3, \quad (77)$$

where the superscript 'd' on the alphas denote that the inequalities in (77) apply to diamagnetic relative permeability. It should be pointed out that the relative permeability (like the relative permittivity) can have values of $\alpha_{ll} < 1$ and even $<0$ without being diamagnetic, and thus these nondiamagnetic $\alpha_{ll}$ would obey the inequalities in (74) rather than (77). For example, metamaterials consisting of arrays of resonant magnetodielectric spheres can have values of relative permeability and permittivity considerably less than zero [39, figures 26 and 28], [40].

*6.2. Implications of causality-energy conditions for static fields*

Letting $\omega \to 0$ in the inequalities of (74) and (77) and assuming that $\lim_{\omega \to 0} \omega d\alpha_{ll}/d\omega = 0$ and $\lim_{\omega \to 0} \omega d\alpha_{ll}^d/d\omega = 0$, we find in addition to $d\alpha_{ll}/d\omega \geqslant 0$ and $d\alpha_{ll}^d/d\omega \geqslant 0$

$$\alpha_{ll} - 1 \geqslant 0, \quad \omega \to 0 \quad (78)$$

and

$$\alpha_{ll}^d (1 - \alpha_{ll}^d) \geqslant 0, \quad \omega \to 0 \quad (79a)$$

or, equivalently

$$0 \leqslant \alpha_{ll}^d \leqslant 1, \quad \omega \to 0. \quad (79b)$$

The inequality in (78) implies that the diagonal elements of the static relative permittivity must be greater than unity. However, the equations in sections 5.1 and 5.2 applied to an electrostatic cloak rather than a magnetostatic cloak show that the radial permittivity element for an electrostatic cloak would have to take on values as small as zero. Moreover, without the assumption that $\lim_{\omega \to 0} \omega d\alpha_{ll}/d\omega = 0$, the inequalities in (74) imply that either the inequality in (78) holds or $\alpha_{ll} \to -\infty$ as $\omega \to 0$. Thus, the spherical and cylindrical electrostatic cloaks corresponding to the magnetostatic cloaks are impossible to produce because they require permittivities that violate the fundamental restrictions of causality and/or energy conservation.

In contrast, the inequalities in (79b) allow for diamagnetic relative pemeabilities that go to zero and thus there appears no fundamental reason that the spherical and cylindrical magnetostatic cloaks derived in sections 5.1 and 5.2 could not be produced. Shore and Yaghjian [39, figure 8] have shown that three-dimensional periodic arrays of closely packed perfectly electrically conducting spheres exhibit approximately isotropic relative permeabilities with values between 0.4 and 1.0 as $\omega \to 0$. More recently, Magnus *et al* [41, 42] have produced a superconducting anisotropic static metamaterial with the relative permeability in one direction having a value that is much less than unity.

Finally, we note that the conditions we have obtained from the inequalities in (74) as $w \to 0$ also imply that the $\omega > 0$ cloaking given in sections 2 and 3, like the electrostatic cloaking, is not realizable even at a single frequency if the frequency is low enough (because of the required zero and near zero values of the radial relative permittivity). The 'low enough' frequency range refers to the quasi-static regime of the individual dipole scatterers that comprise the material or metamaterial.

## 7. Conclusion

Solving the one homogeneous first-order Maxwell differential equation in (15) holding for all possible incident fields with $\omega > 0$ and satisfying the boundary conditions in (16) is shown to be sufficient to derive the lossless permittivity and permeability dyadics as well as the fields inside an ideal electromagnetic annular cloak, which has zero scattered fields outside the cloak and zero total fields within the inner cavity of the cloak, in terms of a general compressed coordinate function $\mathbf{f}(\mathbf{r})$. It is shown that the relative permittivity and permeability dyadics of any shaped annular cloak must be identical and that the tangential components of the **E** and **H** fields as well as the normal components of the **B** and **D** fields must be continuous across the outer surface of the cloak. The normal components of the **B** and **D** fields are required to be zero at the inner material surface of the cloak. For the circular cylindrical cloak, these zero normal-component boundary values do not lead to all the tangential components of the **E** and **H** fields being continuous across the inner surface of the cloak—and thus delta functions arise in the tangential components of the **B** and **D** fields, respectively, at the inner surface of such cloaks [6]. Since homogeneous cavity modes can exist within the free-space interior region of any cloak, it is shown that these homogeneous solutions inside the free-space cavity of the cloak are uncoupled from the fields outside the free-space cavity that are produced by sources external to the cloak and thus the cavity modes are not excited by external sources.

We find that spherical and circular cylindrical cloaks of inner radius $a$ and outer radius $b$ can be formed with a general coordinate function obeying the boundary conditions $f(b) = b$ and $f(a) = 0$, and that the Pendry *et al* spherical cloak [1] and circular cylindrical cloak [11] are particular examples of these spherical and circular cylindrical cloaks. For $H$-wave incident fields, a nonmagnetic circular cylindrical annulus is found that has nonzero scattered fields but still zero total fields within its interior cavity.

If the inner radius of the spherical or cylindrical cloak is zero ($a = 0$), the formulation produces perfectly nonscattering bodies that are not limited to electrical sizes within the dipolar or lower order multipolar regimes. Properly choosing the radial coordinate function yields nonscattering spherical and circular cylindrical concentrators that magnify the incident fields near their centers—one of which corresponds to the cylindrical concentrator in [33].

Except for the arbitrary incident fields and possibly the coordinate function $\mathbf{f}(\mathbf{r})$, the equation (15) is independent of frequency. Therefore, if the coordinate function $\mathbf{f}(\mathbf{r})$ is chosen independent of frequency, the associated relative permeability–permittivity dyadic is independent of frequency. This frequency independence can hold only approximately over a nonzero bandwidth, however, because of the restrictions imposed by causality and energy conservation for lossless materials. Thus, the cloaking of realistic time-dependent incident fields must be approximate.

The boundary value formulation for time-harmonic fields with $\omega > 0$ is appropriately modified to obtain cloaks for static fields ($\omega = 0$). However, causality-energy conditions prohibit the values of the relative permittivity required for electrostatic cloaking and thus only magnetostatic cloaks are realizable. Moreover, the causality-energy conditions imply that even single-frequency cloaking (using the permittivity and permeability dyadics derived herein), in the quasi-static frequency range of the individual dipole scatterers comprising the material or metamaterial, is not realizable except for magnetostatic cloaking at $\omega = 0$.

Lastly, we mention that the possibility of hiding an electrically large, geometrically complicated body with a cloaking layer that is a fraction of a wavelength thick appears (to us) highly unlikely in the foreseeable future even at microwave frequencies. Cloaking with such thin cloaking layers on electrically large bodies would require some of the elements of the identical anisotropic relative permittivities and permeabilities to have very high values and some to have near zero values across the cloaking layer. Even if metamaterials could be developed in the future to approximate these extreme values, they would likely be too lossy to keep the scattering low in the forward hemisphere. Moreover, causality would preclude the possibility of maintaining both the low loss and the bandwidth necessary for reasonably good cloaking in the presence of pulsed incident fields interrogating electrically large bodies.

## Acknowledgments

We are grateful for the helpful insights of Dr John M Myers of Harvard University on the uniqueness of solution for cloaking, and for the support of this work through Dr Arje Nachman of the US Air Force Office of Scientific Research (AFOSR).